\documentclass[]{spie}
\usepackage[]{graphicx}
\usepackage{amsmath}

\title{The GRAVITY metrology system:\\modeling a metrology in optical fibers}

\author{N. Blind\supit{a},  H. Huber\supit{a}, F. Eisenhauer\supit{a}, J. Weber\supit{a}, S. Gillessen\supit{a}, M. Lippa\supit{a}, L. Burtscher\supit{a}, O. Hans\supit{a}, M. Haug\supit{a}, F. Haussmann\supit{a}, S. Huber\supit{a}, A. Janssen\supit{a}, S. Kellner\supit{a}, Y. Kok\supit{a}, T. Ott\supit{a}, O. Pfuhl\supit{a}, E. Sturm\supit{a}, E. Wieprecht\supit{a}, A. Amorim\supit{b}, W. Brandner\supit{c}, G. Perrin\supit{d}, K. Perraut\supit{e}, C. Straubmeier\supit{f}
\skiplinehalf
\supit{a} Max-Planck-Institut f\"ur extraterrestrische Physik, 85748 Garching, Germany; \\
\supit{b} SIM, Fac. de Ci\^encias da Univ. de Lisboa, Campo Grande, Edif. C1, P-1749-016 Lisbon, Portugal;\\
\supit{c} Max-Planck-Institut f\"ur Astronomie, K\"onigstuhl 17, 69117 Heidelberg, Germany;\\
\supit{d} LESIA, Observ. de Paris Meudon, 5, place Jules Janssen, 92195 Meudon Cedex, France;\\
\supit{e} Institut de Plan\'etologie et d'Astrophysique de Grenoble (IPAG) UMR 5274, UJF-Grenoble 1/CNRS-INSU, Grenoble, France;\\
\supit{f} I. Physikalisches Institut, Universit\"at zu K\"oln, Z\"ulpicher Strasse 77, 50937 K\"oln, Germany.\\
}

\authorinfo{N. Blind: E-mail: blind@mpe.mpg.de}

\begin{document}
  \maketitle 

\begin{abstract}
GRAVITY is the second generation VLT Interferometer (VLTI) instrument for high-precision narrow-angle astrometry and phase-referenced interferometric imaging. The laser metrology system of GRAVITY is at the heart of its astrometric mode, which must measure the distance of 2 stars with a precision of 10 micro-arcseconds. This means the metrology has to measure the optical path difference between the two beam combiners of GRAVITY to a level of 5\:nm. The metrology design presents some non-common paths that have consequently to be stable at a level of 1\:nm. Otherwise they would impact the performance of GRAVITY. The various tests we made in the past on the prototype give us hints on the components responsible for this error, and on their respective contribution to the total error. It is however difficult to assess their exact origin from only OPD measurements, and therefore, to propose a solution to this problem.

In this paper, we present the results of a semi-empirical modeling of the fibered metrology system, relying on theoretical basis, as well as on characterisations of key components. The modeling of the metrology system regarding various effects, e.g., temperature, waveguide heating or mechanical stress, will help us to understand how the metrology behave. The goals of this modeling are to {\it 1)} model the test set-ups and reproduce the measurements (as a validation of the modeling), {\it 2)} determine the origin of the non-common path errors, and {\it 3)} propose modifications to the current metrology design to reach the required 1nm stability.
\end{abstract}




\section{Introduction}

GRAVITY is the second generation VLT Interferometer (VLTI) instrument for high-precision narrow-angle astrometry and phase-referenced interferometric imaging\cite{eisenhauer_2011a}.. Its main goal is to observe highly relativistic motions of matter close to the event horizon of Sgr A*, the massive black hole at center of the Milky Way. To do so, GRAVITY will combine four telescopes of the VLTI, and will be assisted by four adaptive optics, plus a near-infrared fringe-tracker. During a typical observation, it will provide simultaneous interferometry of two objects within its 2'' field of view, and shall allow measurements of angles of order 10$\mu as$. To this end, GRAVITY also includes a dedicated metrology system, the design of which was presented in Gillessen (2012)\cite{gillessen_2012a}. This 10 $\mu as$ astrometric accuracy on sky, translates into a measurement accuracy of the optical path difference (OPD) of 5\:nm at maximum. According to the metrology error budget\cite{gillessen_2012a}, the metrology non-common paths (NCP) must be more stable than 1\:nm over at least 1 hour.

This very demanding constraint was addressed in the laboratory by means of a mock-up of the metrology injection unit close to its design. This test set-up was thoroughly used, but its performance were limited by several intrinsic issues (e.g. camera fringing), that prevented us to fully validate the design and understand the origin of the observed NCP error till a year ago. A new Fizeau set-up solved most of these problems (see Sect.~\ref{part:testsetup} for details).

We present in this paper a work that aimed at modelling the metrology injection unit, to understand the origin of the NCP errors and then to look for improvement of the current design. In Section~\ref{part:testsetup}, we present the test set-up, and a summary of the different measurements. In Section~\ref{part:heating_fiber_IO}, we report on measurement and simulations concerning heating in the metrology components, that could be at the origin of the NCP error. Section~\ref{part:met_modeling} presents the modeling of the different test set-up components (i.e. metrology components and camera). Section~\ref{part:test_modeling} aims at validating the model by reproducing the lab measurements, before analysing the full system in Section~\ref{part:fullmet_modeling}, and concluding in Section~\ref{part:ccl}.

\section{Test set-up}
\label{part:testsetup}

\begin{figure}[b]
\centering
\includegraphics[width=0.48\textwidth, trim=0 0 0 3cm, clip=true]{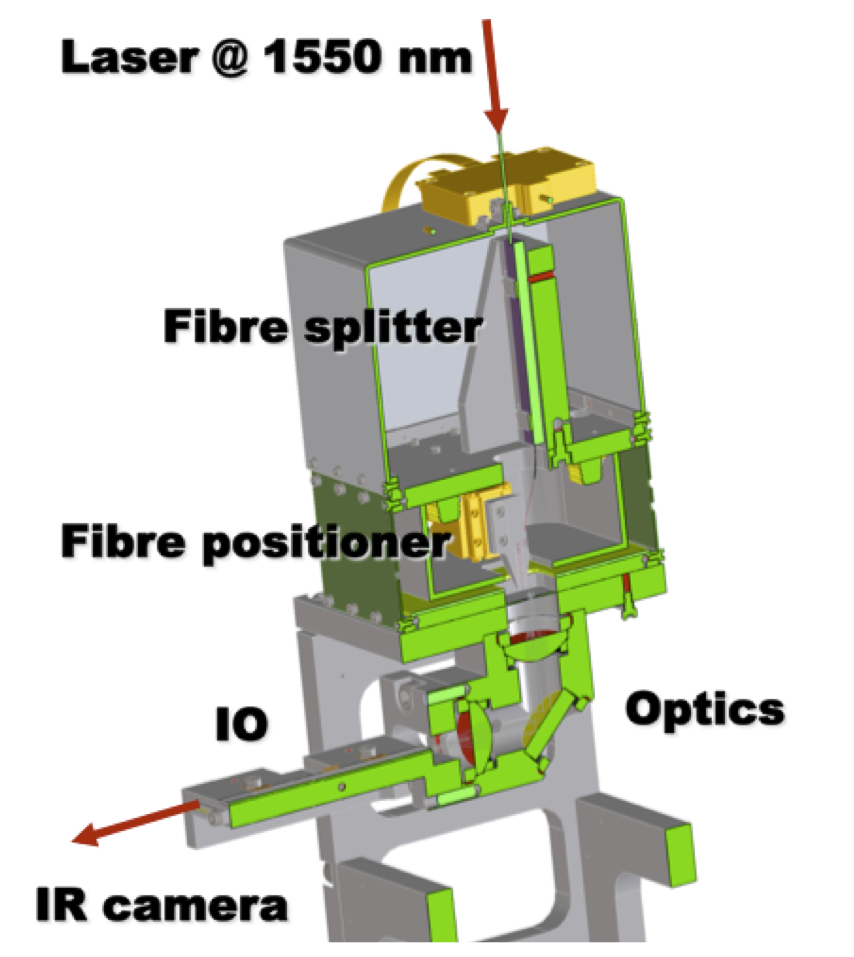}
\includegraphics[width=0.48\textwidth, trim=6cm 3cm 0 3cm, clip=true]{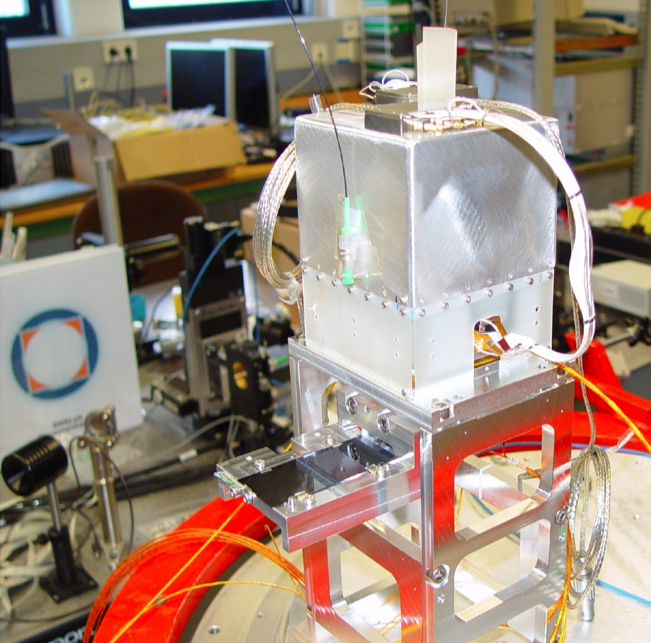}
\caption{Left: Cut-view drawing of the metrology test set-up with the fiber splitter (referred to as {\it exit splitter} hereafter). Right: Picture of the actual set-up before closing the cryostat, with the IOBC on the left bottom part.} \label{fig:metrology-mockup}
\end{figure}

\begin{figure} \centering
\includegraphics[width=0.8\textwidth]{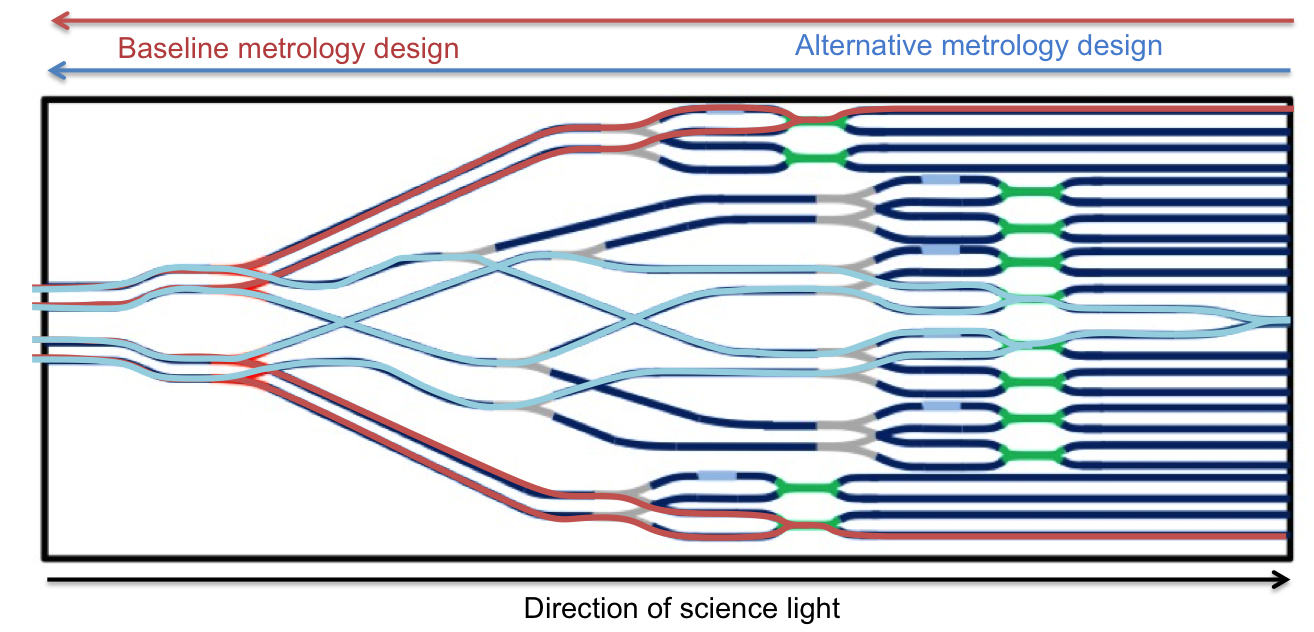}
\caption{Schematics of the GRAVITY alternative IOBC. For the baseline, the Y-junction is absent of the design, which then presents 24 useful outputs. The metrology enters in IOBC from the right side, and feed the four telescopes on the left side. The baseline metrology path is represented in red, while the alternative one is in light blue.} \label{fig:IOBC}
\end{figure}

In order to test the injection schemes, we have built a mock-up of the metrology injection unit (see Fig.~\ref{fig:metrology-mockup}). The mock-up is contained in a small test cryostat, allowing operations in vacuum, but not in cold conditions like in GRAVITY. The metrology is going through the Integrated Optics Beam Combiner\cite{jocou_2012a}(IOBC) that will combine the light from the four VLTI telescopes. Two different integrated beam combiners are considered to inject the metrology light into GRAVITY (Fig.~\ref{fig:IOBC}):
\begin{itemize}
	\item  The baseline design consists of a "classical" component providing a full ABCD fringe coding for each of the six baselines (24 outputs): 2 outputs are used to inject the metrology backward into GRAVITY.
	\item A variant of the previous design (so-called {\it alternative} design), in which the two central outputs are combined by means of a Y-junction, then used to inject the metrology with a single fiber, rather than with an external splitter. In this configuration the two central outputs are lost for the stellar light analysis.
\end{itemize}

\noindent In the test set-up (Fig.~\ref{fig:metrology-mockup}), the laser light is injected via a fiber, and first passes a linear polariser before it reaches an integrated fiber splitter (referred to as {\it exit splitter} hereafter), to which two short exit fibers are glued. The fiber ends are without connectors and each glued into a V-groove, mounted on two SmarAct actuators for XYZ positioning. An optics images the fiber exits onto the IOBC. Each fiber hits a different output for the baseline design, and only one hits the Y-junction in the alternative design (the other fiber misses the IOBC and is therefore not coupled. Between the output lens and the IOBC, a linear polariser is mounted, the orientation of which can be adjusted to the IOBC. Because heating was expected from the high laser power, thermal sensors and thermistors were implemented close to the most critical components, the splitter and the IOBC. Their temperature is then controlled to $\pm$\:3\:mK in practice.

In the first version of the test, the IOBC was used in the "science light" direction to interferometrically combine the light and analyse the phase between the two beams. We could analyze the OPD variations by calculating the phases from a set of ABCD outputs. Hence, the test was not exactly covering the same optical paths as in the instrument later, and one might ask whether there was systematic differences between light traveling forward or backward in the IOBC. Nevertheless, the IOBC being then a phase analyser, the splitter and bulk optic parts are still realistically tested. We later discovered that this set-up was suffering from an important fringing of the infrared camera. It typically limited the measurements precision to $\pm$ 10\:nm (see Sect.~\ref{part:camera_fringing}), and could not demonstrate the metrology injection unit was reaching the requirements.

Since then, we modified the set-up to a Fizeau like one: the IOBC is now mounted in the same orientation as in GRAVITY. The metrology light is injected from the science outputs, which propagates till the four inputs like in Fig.~\ref{fig:IOBC}. Since outputs are in a redundant configuration, one of them is blocked, and the three others are directly projected onto the IR camera detector plane, where they interfere at three different spatial frequencies. By orienting properly the camera and using an appropriate Fourier transform analysis, it was then possible to disentangle the metrology fringe signal from the detector fringing. This allowed to reach far more consistent results.

\section{Heating in IOBC and fibers}
\label{part:heating_fiber_IO}

\subsection{Results from test set-ups}

\begin{figure} \centering
\includegraphics[width=0.9\textwidth]{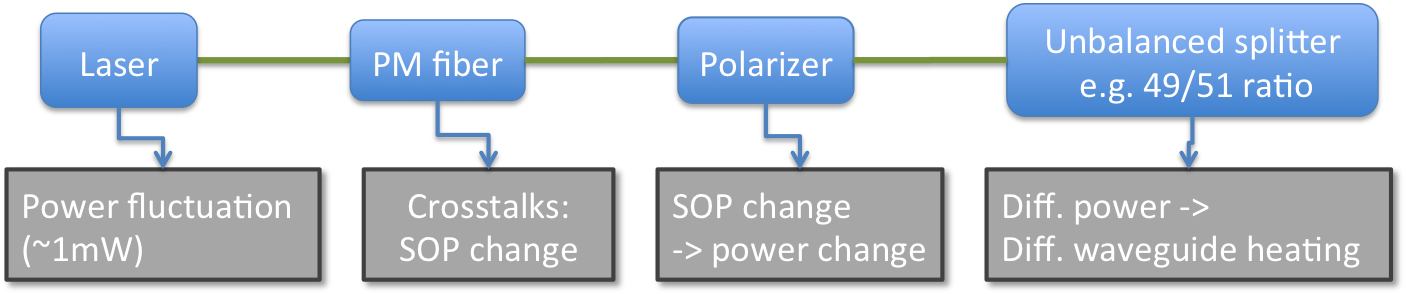}
\caption{Schematics representation of the heating chain in the metrology.} \label{fig:metrology_heating}
\end{figure}

The temperature control of the metrology is critical to ensure its stability to a nanometer level, regarding its mechanical and optical properties. Although the cryostat and the different components are stabilized to better than 10\:mK, various test showed still a strong correlation between the OPD instability and both the ambient temperature and the laser power. A laser power sensitivity was measured down to an initial power of 50\:mW (less than 10\:mW then reaching the critical components). The power senstivity was also measured a factor of 20 higher in vacuum than in air, strongly suggesting a less efficient component cooling in vacuum via radiation only  (see Tab. \ref{tab:power_sensitivity}). 
\begin{table}[b] \centering
\begin{tabular}{c|cc|c}
\hline \hline
& \multicolumn{2}{c|}{Baseline} & alternative \\
& Splitters & IOBC & IOBC \\
\hline
Ambient & 0.25 nm/mW& 0.25 nm/mW& 0.25 nm/mW\\
Vacuum & 5.0 nm/mW& 0.25 nm/mW& 0.25 nm/mW\\
\hline
\multicolumn{4}{c}{}
\end{tabular}
\caption{Estimated power sensitivity of the IOBC and splitter in the initial test set-up, under atmospheric pressure and in vacuum.}\label{tab:power_sensitivity}
\end{table}
A possible scenario is that the laser power fluctuations (measured at $\sim$1\:mW level) translate into variable, differential heating in the metrology chain, as shown on Fig.~\ref{fig:metrology_heating}. The external Polarisation Maintaining (PM) fiber and the following entrance polariser could also generate differential heating: because of polarisation crosstalks in the PM fiber, the output State Of Polarisation (SOP) changes with temperature, which then turns into power fluctuations when passing the entrance polariser. According to measurements, the relative power fluctuations is $\sim 2 \cdot 10^{-3}$ for the 2\:m long PM fiber used for the tests, and up to $\sim 5\cdot10^{-2}$ for the 30\:m long fiber that should be used in Paranal (see Sect.~\ref{part:PMfiber}). Following this scenario, differential heating in the fiber components would be the main instability contributor. An initial step of the metrology modeling is then to realistically determine the level of heating reached in the central waveguides with respect to the injected laser power.

     \subsection{IOBC heating}
		\subsubsection{Observations}

\begin{figure} \centering
\includegraphics[width=0.43\textwidth]{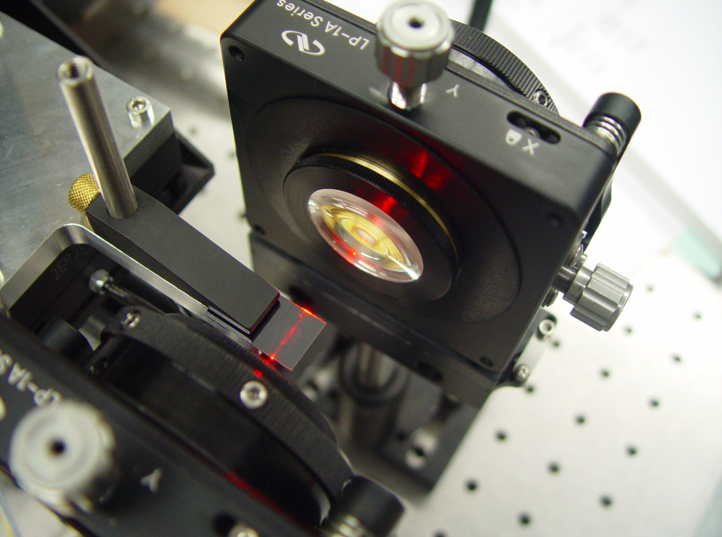} \hfil
\includegraphics[width=0.49\textwidth]{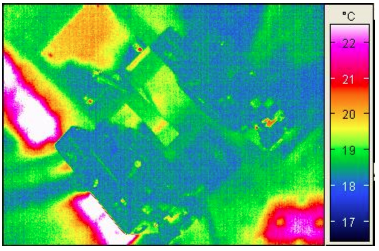}
\caption{Left: picture of the transmission and heating measurement set-up, where an IR laser is injected into the central IOBC. The red alignment laser is visible here. Right: thermal picture of the set-up after 5 minutes, and 0.3 W of laser power. The component is heated by $\sim$ 1K.} \label{fig:IO_BARTKO}
\end{figure}

Direct lab measurements showed a significant heating of the IOBC surface under atmospheric pressure, up to several 10\:K (Fig.~\ref{fig:IO_BARTKO}). Since the exit splitters absorption is also relatively important ($\sim$ 60\% transmission for a few cm length) we can expect a similar heating. These critical components were actively temperature controlled via different heaters to face this issue. Despite this thermal control, we ignore the actual temperature reached at the waveguide core level, buried 30\:microns under the surface, and only 5-6\:microns large (to compare to the cm scale of the surface). The metrology being in addition operated (mostly) in vacuum, heating could also be far more important since convection is lost as dissipation mechanism.

According to the FDR document of the IOBC:
\begin{itemize}
     \item The transmission of a prototype IOBC (including all functions needed to combine and analyse four telescopes) was measured to 45\%;
     \item The transmission of a straight waveguide from the same IOBC wafer was measured to 55\%.
\end{itemize}
Both components having the same length (8\:cm), we can consider that 45\% of the metrology light is really absorbed by the IOBC components, while 10\% are lost as stray light in the substrate layer through the different optical functions. The measurements presented in Fig.~\ref{fig:IO_BARTKO} led to a 1 to 3\:K uniform surface heating when injecting 0.3\:W laser power, and 5 to 30\:K for 1.9\:W laser power. We roughly estimate the heating efficiency to be in the range 5-15\:K/W. Having only 3 such measurements, all in different conditions, different laser power, etc. it is difficult to give a more precise result. An additional measurement showed that when hitting directly the layer (and not a waveguide) no heating was generated.

The question of heating in vacuum is then still open, which we addressed with simulations with the software COMSOL (first under ambient pressure to validate the model, and then in vacuum).

               \subsubsection{Simulation at ambient pressure}

\begin{figure} \centering
\begin{minipage}[t]{0.44\textwidth}
\includegraphics[width=0.99\textwidth]{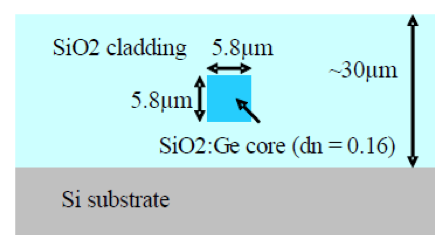}
\caption{Design and dimension of the IOBC. The Si substrate is 750 $\mu$m thick.}\label{fig:IOBC_layer}
\end{minipage}
\hfill
\begin{minipage}[t]{0.54\textwidth}
\centering
\includegraphics[width=0.9\textwidth]{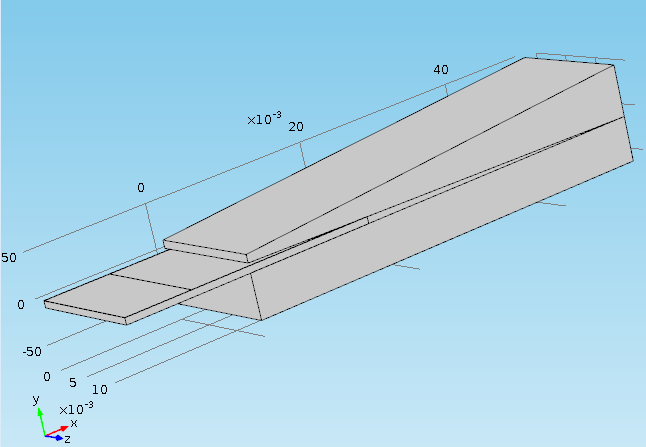}
\caption{COMSOL geometry used to reproduce the lab heating measurements.}\label{fig:IO_heating_lab}
\end{minipage}
\end{figure}

As a first step, we modelled with the software COMSOL Multiphysics the laboratory set-up consisting of an IOBC and a plier. Fig.~\ref{fig:IO_heating_lab} shows the simplified model, where:
\begin{itemize}
	\item The IOBC consists of a 5.8\:$\mu$m large Ge doped silica waveguide, burried in a 30\:$\mu$m thick silica layer, on a 750\:$\mu$m silicon substrate (Fig.~\ref{fig:IOBC_layer}). The IO chip is 8\:cm long, and 5\:cm large, and its transmission was estimated to 70\%.
	\item Laser power is dissipated in the IOBC along a square section tube with the dimension of the waveguide (with silica thermal properties).
   \item The mount is made of standard aluminum.
\end{itemize}
We applied the following boundary conditions:
\begin{itemize}
     \item A constant room temperature on the back face of the mount (T=293\:K at x=50\:mm).
     \item Cooling by natural air convection on the other faces, with surface thermal conductivity of 1, 3 and 4 $W.m^{-2}.K^{-1}$ for the bottom, vertical and top faces respectively\cite{rowley_1930a}.
     \item Surface radiation was also considered, by setting emissivity of 0.9 for the IOBC (silica glass), and 0.8 for the black anodized aluminum mount.
\end{itemize}
For the different measurements, we could reproduce the order of magnitude of IOBC heating:
\begin{itemize}
     \item For P=1.9\:W, the surface temperature increases by $\sim$ 15\:K, and we observe a gradient of 6\:K between the IOBC surface and the waveguide;
     \item For P=0.3\:W, the surface temperature increases by $\sim$ 7\:K, and the gradient to the waveguide reaches 2-3\:K.
\end{itemize}

     \subsection{ Simulations of IOBC in vacuum}

\begin{figure} \centering
\includegraphics[width=0.49\textwidth, trim=0cm 2.cm 0 2.5cm, clip=true]{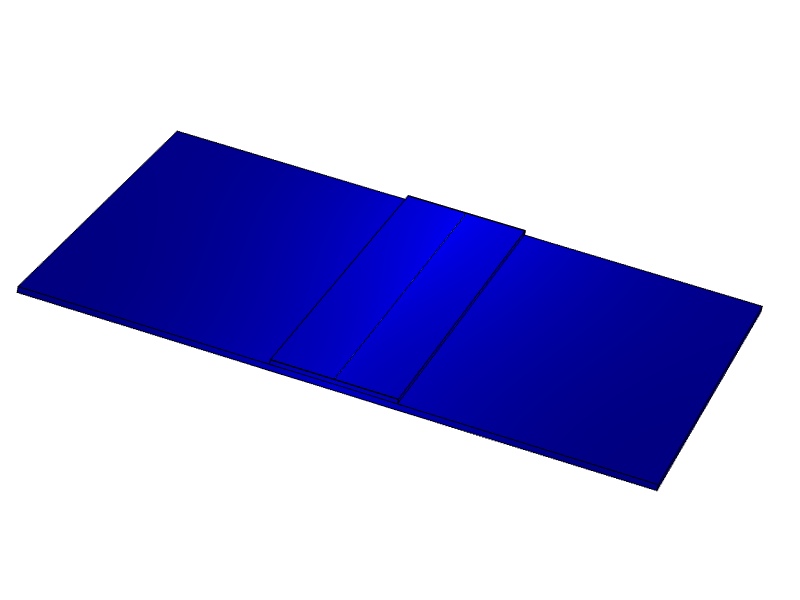}
\caption{COMSOL model of the IOBC mounted on an aluminium plate.} \label{fig:IO_heating_vacuum_model}
\end{figure}

\begin{figure} \centering
\includegraphics[width=0.49\textwidth]{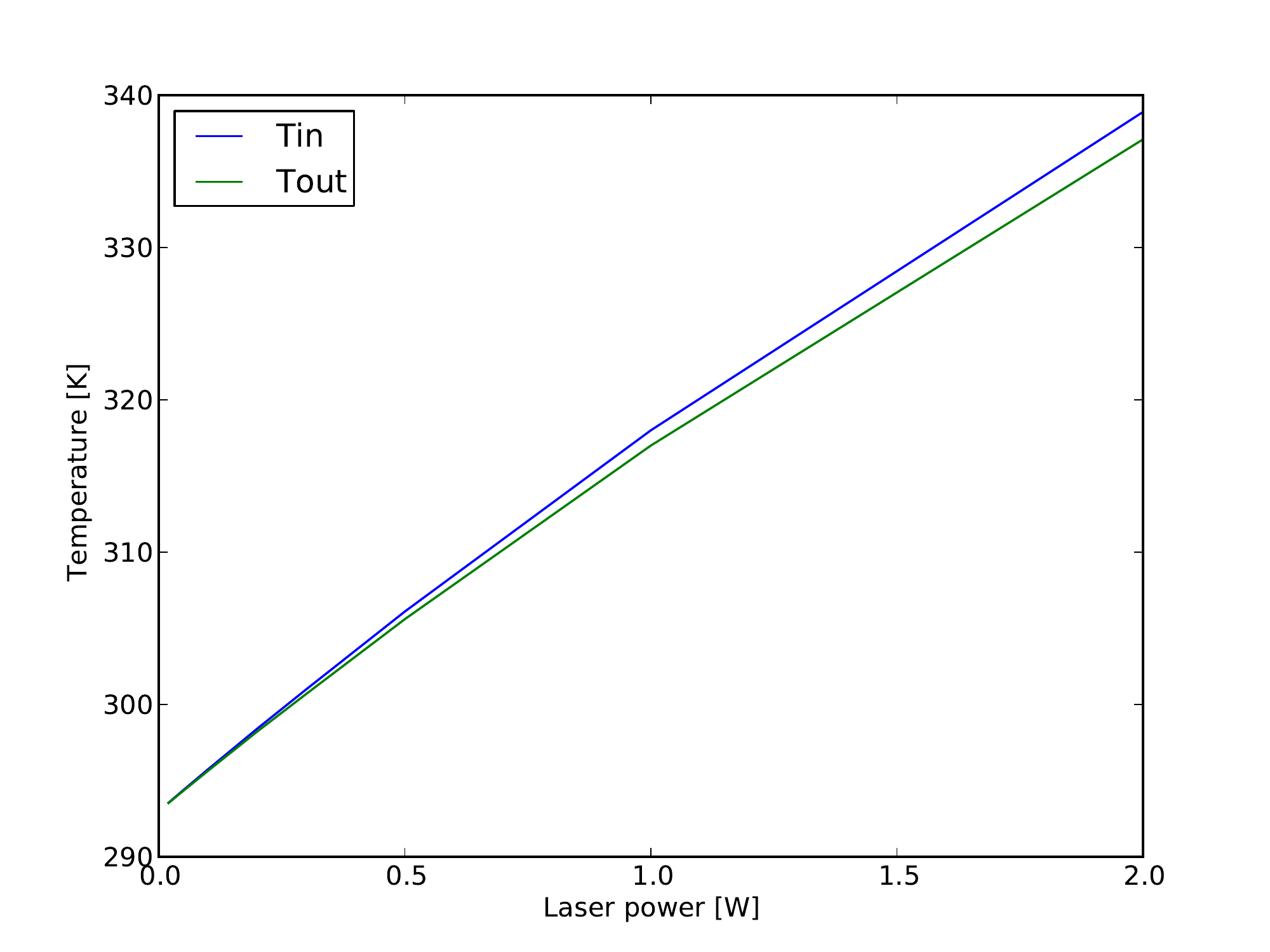}
\includegraphics[width=0.49\textwidth]{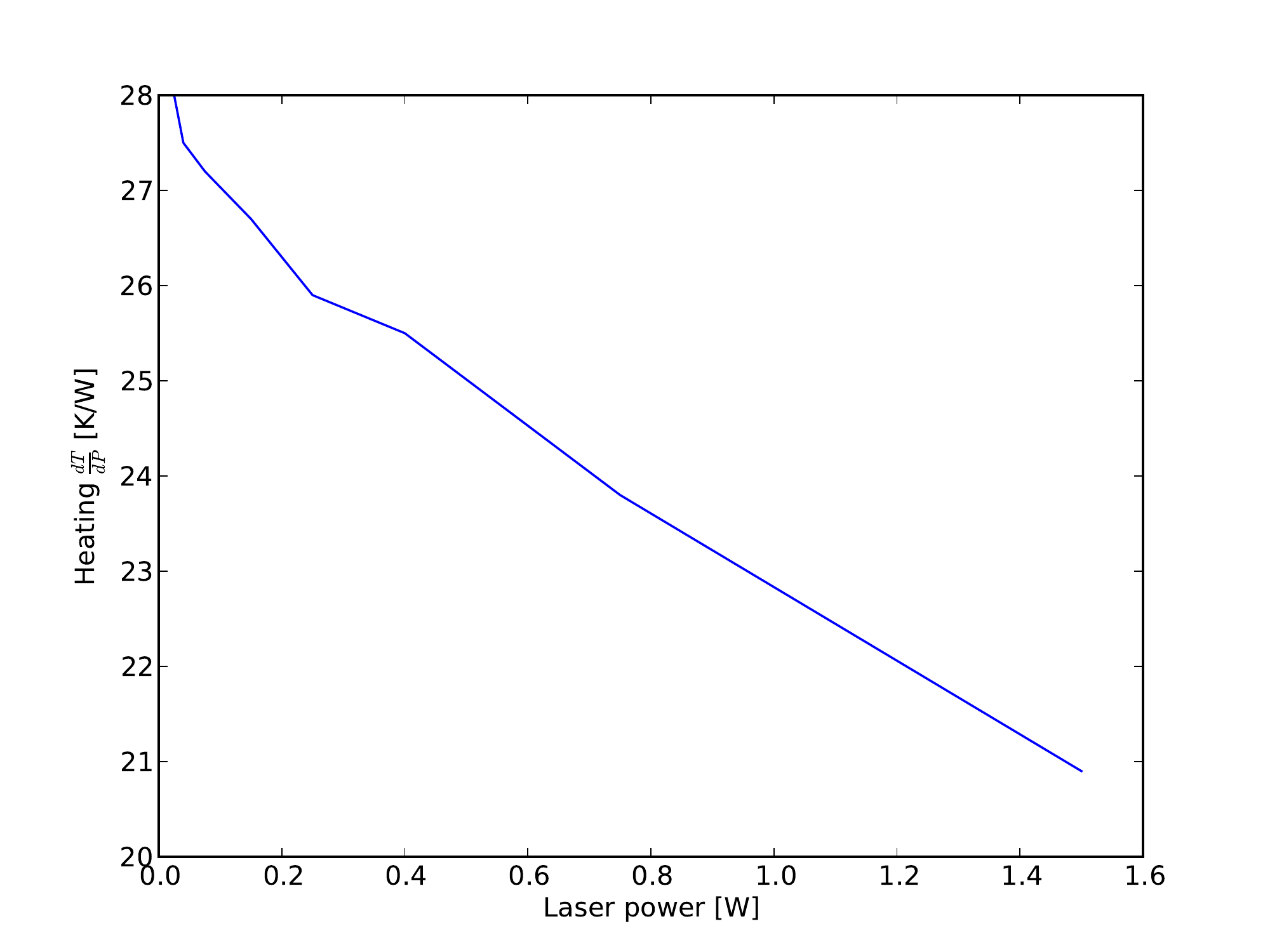}
\caption{Left: temperature increase in the IOBC waveguide as function of the injected power, at the entrance (blue) and output (green). Right: waveguide heating as function of the power.}\label{fig:IO_heating_vacuum}
\end{figure}

We used again COMSOL to estimate the IOBC heating in vacuum, corresponding to the test and instrument conditions. We considered the final 54\:mm long component with a transmission of 70\%, that we modeled the same way as in the previous section, and placed it on a large aluminum plate, representing the support mount in the test set-up (Fig.~\ref{fig:IO_heating_vacuum_model}). The boundary condition was a constant ambient temperature (T = 293\:K). We accounted for thermal cooling by surface radiation only, and did not take into account the semi-transparent nature of the silica layer. Simulations for different power levels up to 2\:W allowed to extract a power sensitivity at the waveguide level (Fig.~\ref{fig:IO_heating_vacuum}):
\begin{equation}
\frac{d T}{d P} \sim 20 \mathrm{-} 30 \:\mathrm{K/W}
\end{equation}
The temperature gradient along the waveguide also increases at a rate of $\sim 2$\:K/W.

\subsection{Heating in bare fibers}

At the exit splitter fiber level, we could measure a temperature increase in vacuum of 5 to 8 K/W, as well as a differential heating between these fibers of the order of 2.5 to 4 K/W. These values do not represent the actual temperature in the fibers themselves since sensors are placed few centimeters away, and therefore represent lower values. 

%
\begin{figure}[b] \centering
\includegraphics[width=0.7\textwidth]{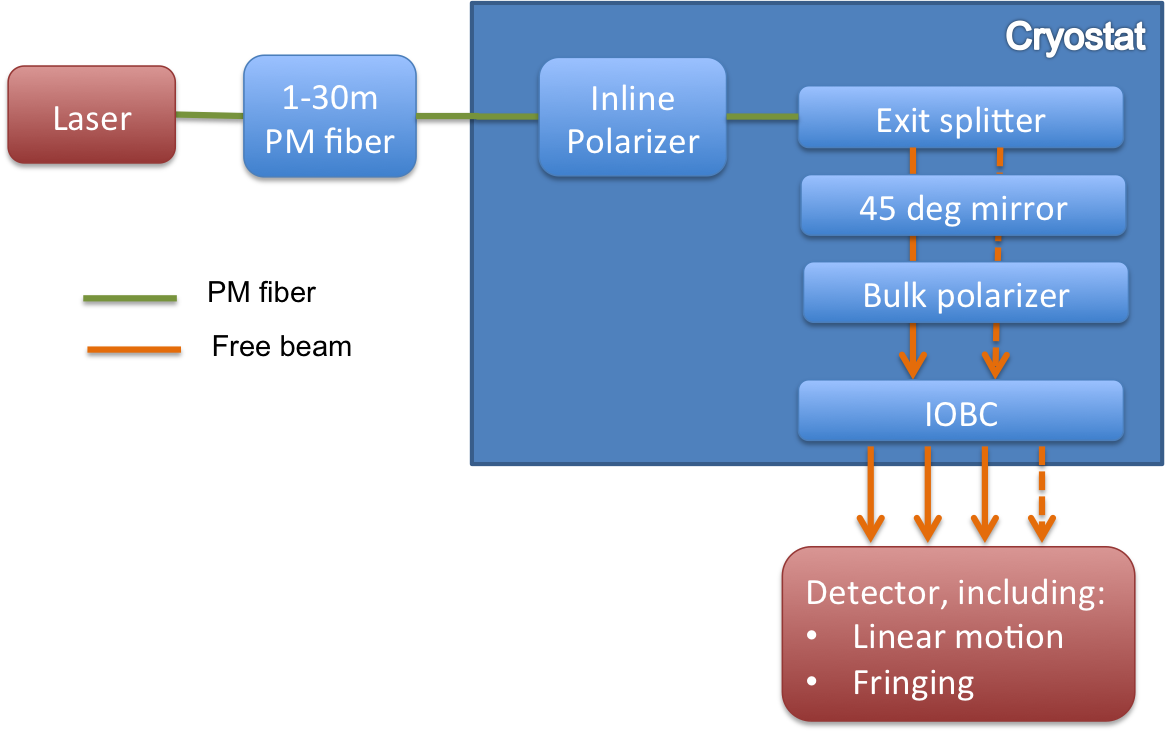}
\caption{Schematic of the metrology injection unit. Depending on the injection deisgn only one exit of the splitter is used, and three or four IOBC outputs are used. The splitter and IOBC are assemblies of sub-functions presented in Sect.~\ref{part:splitter_model} and~\ref{part:IOBC_model}.} \label{fig:injection_schematic}
\end{figure}
%

\section{Injection optics and test cryostat modelling}
\label{part:met_modeling}

The model of the metrology injection unit presented in Sect.~\ref{part:testsetup} (Fig.~\ref{fig:injection_schematic}) consists in propagating the electric field from component to component, each being described by a $2\times 2$ Jones matrix. They are derived either from actual polarisation (when possible) or photometric characterisation of certain components, or from theoretical models. Light power is attenuated from element to element in the metrology chain, according to transmission measurements and data sheet. The energy deposited is used to heat the component according to the previous section results.

The modeling of the lab experiments served as validation. Because of the detector fringing and its position sensitivity (Sect.~\ref{part:camera_fringing}), and because of the important number of free parameters, only few measurements were suitable for such a comparison.

     \subsection{Individual components models}
     
		\subsubsection{Connectors}

The metrology chain includes several standard FC/APC connectors, plus a few splices. They are simply modeled by a rotation matrix at the interface between two components, whose effect is to misalign polarisation axes of one component to the next one:
\begin{equation}
L = \begin{bmatrix}
	\cos(\alpha) & \sin(\alpha) \\
	-\sin(\alpha) &  \cos(\alpha)
	\end{bmatrix}
\end{equation}
, with $\tan \alpha = 10^{-ER/10}$, and ER the extinction ratio in intensity:
\begin{itemize}
   \item ER $\sim$ 15 for FC/APC connector (i.e. misalignment angle of $\pm 2^\circ$);
   \item ER $\sim$ 27 for a fiber splice (i.e. misalignment angle of $\pm 0.1^\circ$).
\end{itemize}
%
%

		\subsubsection{Polarisation Maintaining fiber}
		\label{part:PMfiber}
		
The metrology relies on the use of PM fibers to preserve the linear polarisation of the metrology laser and minimise the impact of birefringent components. They are modeled with the following Jones matrix:
\begin{equation}
L = \begin{bmatrix}
          \sqrt{1 - X^2} \: e^{j \: \phi_x} & X\:  e^{j \: \phi_y} \\
          -X \: e^{j \: \phi_x} & \sqrt{1-X^2} \: e^{j \: \phi_y}
     \end{bmatrix}
\end{equation}
where $X = 10^{-ER/10}$ is the polarisation crosstalk in intensity between the PM fiber axes. $\phi_{x,y} = 2\pi n_{x,y} z/\lambda = k_{x,y} z$ is the absolute phase introduced by the fiber along each of its axis, $z$ being its length. The diagonal terms of the matrix preserve the energy. The extinction ratio decreasing in proportion of the logarithm of the fiber length\cite{fujikura_PANDA}, $X(z) = X_0/ \mathrm{log}_{10} (z/z_0)$, ER$\sim15$ for $z$ = 30\:m, i.e. the fiber between the metrology electronic cabinet and the cryostat. 

\begin{figure} \centering
\includegraphics[width=0.5\textwidth, trim=4cm 4cm 4cm 5cm, clip=true]{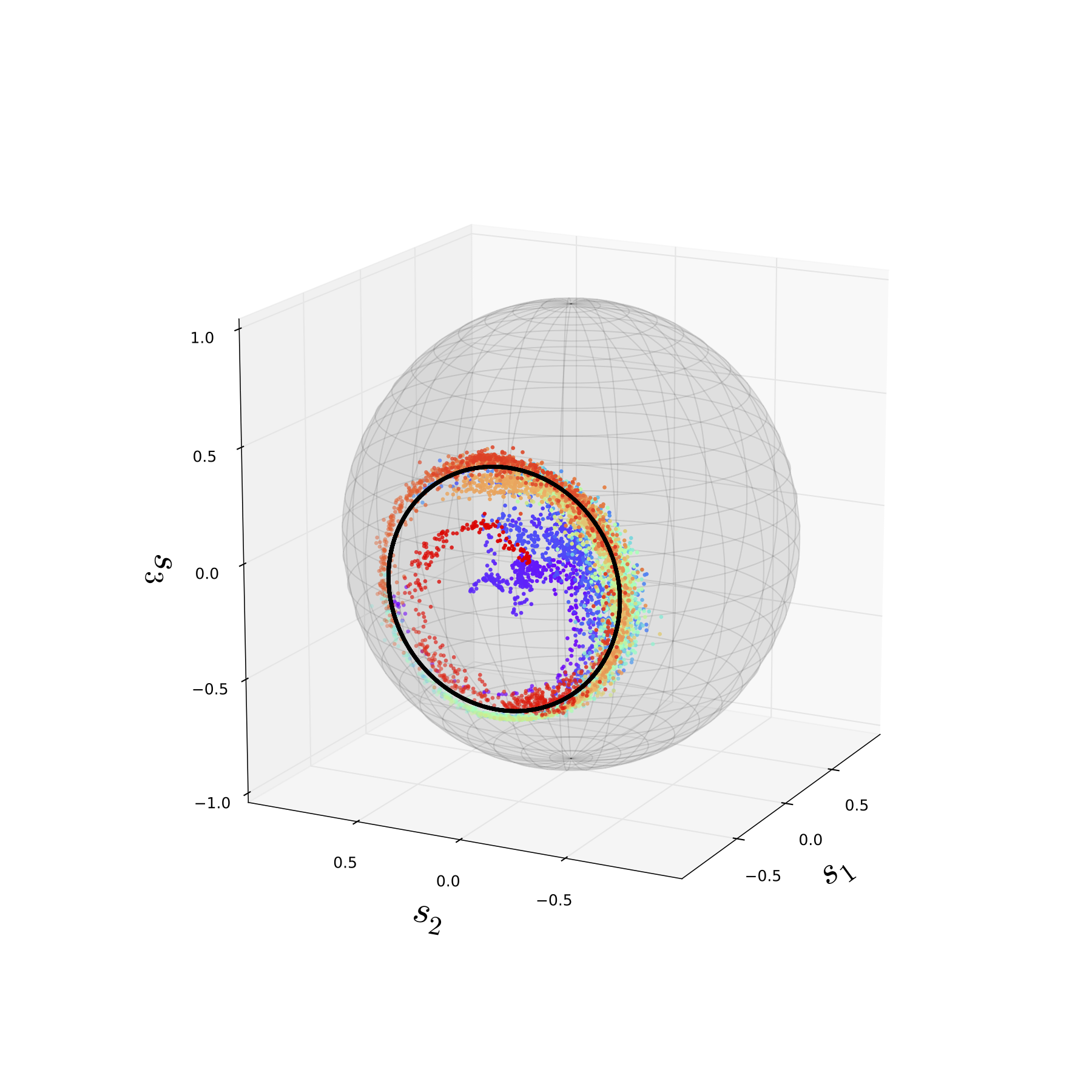}
\caption{Measurement (color) and modeling (black) of the output SOP of our 30\:m long PM fiber (+ metrology laser), projected on the Poincarr\'e sphere.  The color coding represents time. The radius of the circle described by the fiber SOP is directly related to its extinction ratio, and suggests ER$\sim$11. \label{fig:poincarre_PM}}
\end{figure}

We measured the ER of a 80 cm PM fiber around $\sim 23$, as specified by the manufacturer. A measurement during a night with the air conditioning ON ($\Delta T=2\:K$) allowed us to estimate ER$\sim$11 for our 30\:m long fiber (Fig.~\ref{fig:poincarre_PM}). The expected power fluctuation after the entrance polariser is then $\pm$0.5\% and $\pm$3\% of respectively.

         \subsubsection{Exit splitters and SM fibers}
		\label{part:splitter_model}
		
\begin{figure}
\centering
\includegraphics[width=0.8\textwidth]{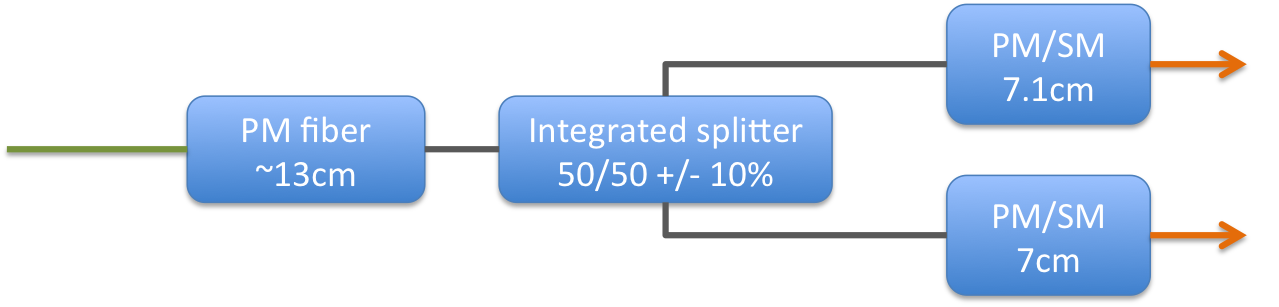}
\caption{Schematic representation of the splitter model. The connections between the different section are splices. The output fibers are not connectorised.} \label{fig:SplitterModel}
\end{figure}

In the baseline design, the metrology light is injected into the IOBC via the exit splitter which consists of (Fig.~\ref{fig:SplitterModel}):
\begin{itemize}
	\item a 13 cm long PM fiber at the entrance; 
   \item an integrated Y-junction with splitting ratio close to 50/50;
   \item  two output fibers, either Single Mode (SM) or Polarisation Maintaining (PM), 5 to 7cm long, equalised to better than 1mm.
\end{itemize}
The splitter is modelled as the series of these 3 components, spliced to each other.
\\

\noindent  {\bf The Y-junction} -- The Y-junction is modelled according to the data sheet provided by the manufacturer, plus a series of measurements. We could observe that their transmission and splitting ratio are independent of the laser power (exit fibers included). A polarimetric characterisation was not possible, as it highly depends on the actual fiber routing, especially for SM fibers.
\\

\noindent {\bf The SM and PM fibers} -- The splitter is the only component of the metrology injection unit potentially having SM fibers. Although PM fibers maintain the polarisation axes, they are far more sensitive to misalignment or twist when they are used for free propagation (i.e. not directly connected to another fiber), the polarisation rotation $\alpha$ following the fiber twist $\theta$:
     \begin{equation}
      \alpha = \theta
     \end{equation}
while for an SM fiber, a pure twist as a factor 12 lower impact on polarisation rotation\cite{ulrich_1979a}:
\begin{equation}
     \alpha = 0.08\: \theta 
\end{equation}
Given the birefringence of the PM fiber ($\Delta n\sim10^{-4}$), we would then expect a polarisation effect on the metrology phase.

\begin{figure} \centering
\includegraphics[width=0.6\textwidth, trim=0cm 0cm 0cm 1cm, clip=true]{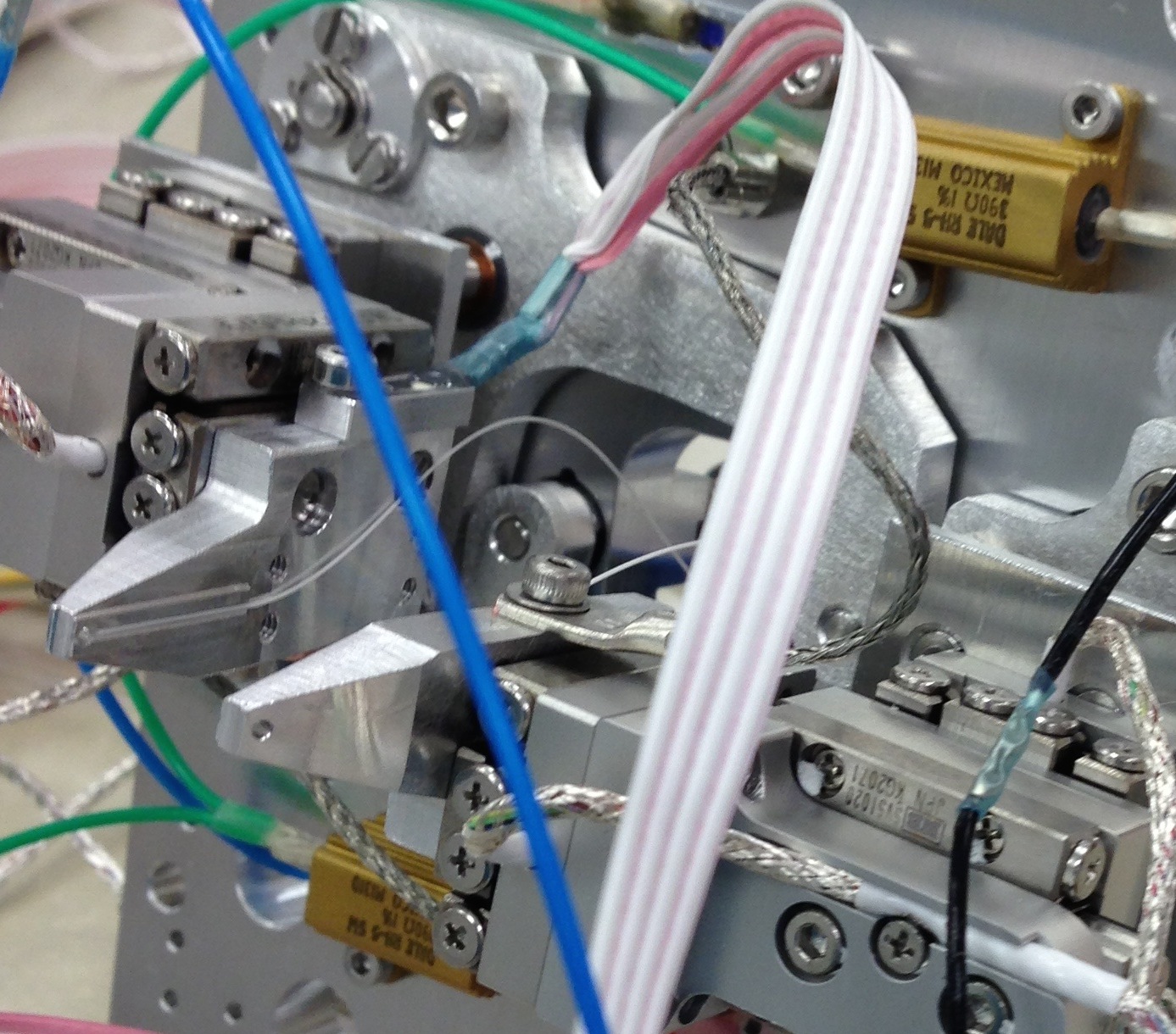}
\caption{Picture of an exit splitter fibers in its V-groove mount. The fibers are bended in a S-like shape, whose plan is not  aligned to the IOBC axes, resulting in a polarisation misalignment.} \label{fig:splitter_fibers}
\end{figure}

We however observed in the lab that the splitter SM fibers behaved like PM ones, i.e. twisting the fiber was rotating the polarisation in a 1-to-1 proportion. While polarisation axes were aligned to better than 2$^\circ$ during their gluing to the V-groove (with straight fibers), misalignment between -40 and +40$^\circ$ were observed once the splitter mounted in the test set-up, where fibers were bended (see Fig.~\ref{fig:splitter_fibers}). Because of the stress induced birefringence, polarisation axes parallel and perpendicular to the bend plane are created. The resulting effect is similar to the one used for Mickey ears polarisation controllers. For fused silica fibers, the SM bending birefringence amounts to:
\begin{equation}
     \beta = -7.7 \cdot 10^7 \kappa^2 r^2\: deg\:.m^{-1}
\end{equation}
at $\lambda=633$ nm\cite{ulrich_1980a, drexler_2012a}, where $\kappa = 1/R$ is the inverse of the bend radius, and $r$ is the fiber cladding diameter ($r = 125 \mu m$ in our case). With an observed bend radius of $\sim$3\:cm, we expect a birefringence of $\Delta n = 2.3 \cdot 10^{-6}$ (i.e. $\Delta \phi = 31^\circ$ for 7cm long fiber at $\lambda = 1908$\:nm). The birefringence is still a factor 100 lower than a typical PM fiber. The main effect is then the uncontrolled misalignment to the IOBC, which is in the range $-$40$^\circ$ to $+$40$^\circ$. On the other hand, the PM fiber were aligned in their V-groove shoes with a precision of $\pm$ 2$^\circ$.

The splitter was consequently moved back into its mount to increase the bend radius of SM fibers. Since we still need some margin to align and focus each fiber to the IOBC, it was not possible to have straight fibers however. It is then not sure that we are free from misalignment to the IOBC.

          \subsubsection{Bulk and inline polarisers}

Polarisers were simply modelled with the following Jones matrix:
\begin{equation}
L = \begin{bmatrix}
        1 & 0 \\[0.3em]
        0 &  10^{-ER/10} \\[0.3em]
      \end{bmatrix}
\end{equation}
with $ER$ the Extinction Ratio in intensity, reaching $\sim$30 dB for bulk components, and 23 dB for in-line polariser (including spliced fibers). The bulk polariser between the exit splitter and the IOBC was aligned to better than a degree to the IOBC axes.

          \subsubsection{Integrated Optics} 
        	\label{part:IOBC_model}
          
The integrated optics was identified as the second contributor to the phase instability of the metrology. Because of the planar nature of the waveguides, the birefringence axes are parallel and perpendicular to the IOBC plane, with a birefringence of $\Delta n \sim 10^{-4}-10^{-3}$. The polarisation axes being by design maintained from the input to the output, birefringence could impact the metrology signal if the splitter exit polarisation is not well aligned to them. The purpose of the bulk polariser placed between the exit splitter fibers and the IOBC is therefore to limit such an effect.

\begin{figure} \centering
\includegraphics[width=0.99\textwidth]{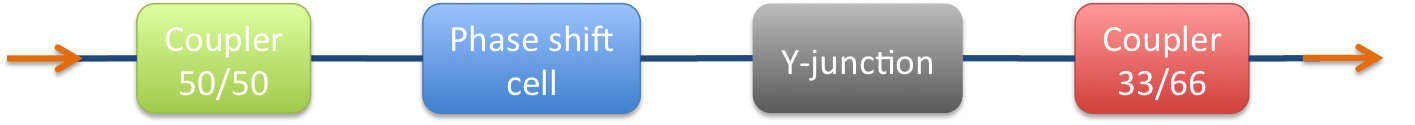}
\caption{The IOBC and metrology path was presented in Fig.~\ref{fig:IOBC} (Sect.~\ref{part:testsetup}). Each arm through which the metrology light propagates can be modeled according to the series of functions schematised here, and linked by straight waveguides.} \label{fig:IOBC2}
\end{figure}

The IOBC was modelled by implementing the different optical functions through which goes the metrology light (Fig~\ref{fig:IOBC2}):
\begin{itemize}
	\item 50/50 coupler (green);
	\item Phase shifting cell (light blue);
	\item Y-junction (in coupling direction, in grey);
   \item 33/66 coupler (in red).    
\end{itemize}
The waveguides crossings in the IOBC were not considered, crosstalks being lower than 1/1000. The characteristics of the IOBC were found in Labeye (2008)\cite{labeye_2008} and Jocou (2012)\cite{jocou_2012a} , including optical properties of silica, birefringence induced by the waveguide technology ($\Delta n \sim 10^{-3}$), plus heating at the level of 10 to 30K/W according to previous simulations.

This IOBC model assumes perfectly SM waveguides. This is a potential caveat of the alternative design model, since the additional Y-junction is slightly multimode, and generates a false phase effect depending on how well light is coupled, of the order of 20\:nm/$\mu$m when moving the fiber across the waveguide. This was not introduced in the model: the SmarAct read back however indicates position variations of amplitude $<$100\:nm, so that this effect could be of the order of $\pm$ 2\:nm. We also do not know how the splitting ratio varies.

The different functions implemented in the model are:
\begin{itemize}
	\item {\bf Waveguide} -- By design, the IOBC waveguides should be close from highly PM, i.e. with very low crosstalk between polarisations. We therefore modelled them simply as perfectly PM:
	\begin{equation}
	L =
		\begin{bmatrix}
			e^{j \:\phi_x} & 0 \\
			0 & e^{j \; \phi_y} \\
		\end{bmatrix}
	\end{equation}
	with $\phi_{x,y} = 2\pi\: n_{x,y}\: z/\lambda = k_{x,y} z$, $z$ being the waveguide length.

   \item {\bf Couplers}  -- Couplers are based on the coupling of the evanescent part of a guided mode into a neighbour waveguide. Depending on the optical path length, it is possible to control the splitting ratio, so as to obtain the 66/33 and 50/50 couplers/splitters of the IOBC\cite{snyder_1983a, payne_1985a, huang_1994a} . For identical waveguides, the output amplitude $E_i^{out}$ of the entrance field $E_1^{in}$ is:
          \begin{eqnarray}
               E_1^{out}(z) /E_1^{in} &=& \cos(\pi/2\: z / L_c) \times e^{-j k z}\\
               E_2^{out}(z)/E_1^{in} &=&  -j \sin(\pi/2 \: z / L_c) \times e^{-j k z}
          \end{eqnarray}
where  $L_c$ is the beat length ($L_c \propto \lambda/n$), governed by the geometry and refractive index profile of the waveguide. 50/50 and 66/33 couplers were characterised and showed departures from ideal values at a level of 1\%.
For $L = 0.4 \: L_c \:\pmod{2L_c}$ and $0.5 \: L_c\:\pmod{L_c}$, we obtain 66/33 and 50/50 couplers respectively. According to Labeye (2008)\cite{labeye_2008}, the coupler length for 66/33 couplers in H-band is $\sim 915 \mu m$, which would linearly translate to $\sim$1100 nm at 1.9\:$\mu m$. Coupler heating induces an expansion of the waveguide, and a decrease of the beat length. Using typical properties of silica and IOBC components, the splitting ratio changes by less than $5\cdot 10^{-5}\:K^{-1}$ and should then have a negligible impact.
   \item {\bf Phase shift cells} -- The phase shifter cells are a series of waveguide sections of varying widths, resulting in sections of slightly different effective indices (Fig.~\ref{fig:PhaseShiftCell}). The index change is of the order of $5\cdot 10^{-4}$ for a waveguide section change of 1 $\mu m$\cite{labeye_2008}, i.e. $\pi/2$ phase shift per millimeter at $\lambda = 1.9 \mu m$. In practice, the total length of the phase shifting cell can reach 1\:cm, as different sections are necessary to limit the chromatism(Fig.~\ref{fig:PhaseShiftCell}). In absence of precise information, they are modeled as straight waveguides.

\begin{figure} \centering
\includegraphics[width=0.7\textwidth, trim=0cm 0cm 0cm 0.1cm, clip=true]{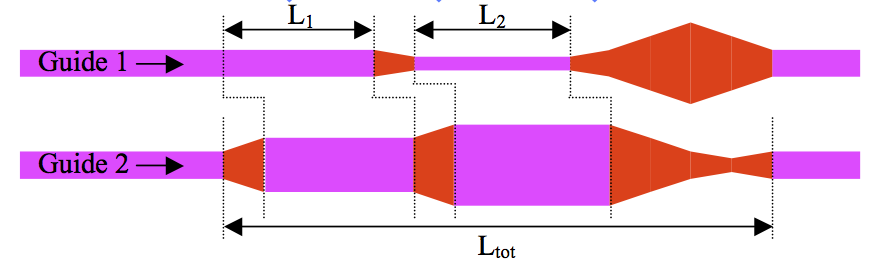}
\caption{Schematic of two parallel waveguides forming an achromatic phase shifting cell made of two sections. The total length is 7.8\:mm.  Taperisation between sections minimises losses. Figure from Labeye (2008)\cite{labeye_2008} .} \label{fig:PhaseShiftCell}
\end{figure}
%

%
%

	\item {\bf The Y-junction splitters} -- The Y-junction was modelled according to the tests of the exit splitters, i.e. the splitting ratio and the transmission are constant for any laser power and temperature, but not necessarily balanced. As they are used here in the reverse direction (i.e. as couplers), $\sim$ 50\% of the light is rejected into the IOBC substrate layer, without generating heating.
\end{itemize}

Because of the symmetry of the design, and the equalisation of the internal optical paths to less than $1 \mu m$, IOBC phase effects may appear because of a wrong {\it differential} injection by the splitter, regarding polarisation (fiber alignment) and splitting ratio (asymmetric heating).

         \subsubsection{Optics}
         
The optical element between the splitter fibers and the IOBC are two lenses and one mirror at 45 degrees inclination in between. We considered perfect lenses, and adopted the following Jones matrix model for the mirror\cite{perraut_1996b, lebouquin_2005a}:
\begin{equation}
     L = \begin{bmatrix}
     1 & 0 \\
     0 &  \tau e^{j\delta\phi}
      \end{bmatrix}
     \end{equation}
where $\tau  = 1 - i^2/10$ and $\delta \phi = - i^2/3$ are the dichroism and retardance of the mirror, and $i$ is its inclination angle in radians.

		\subsubsection{IR camera} \label{part:camera_fringing}

\begin{figure} \centering
\includegraphics[width =0.5\textwidth, trim=0cm 0cm 0cm 1cm, clip=true]{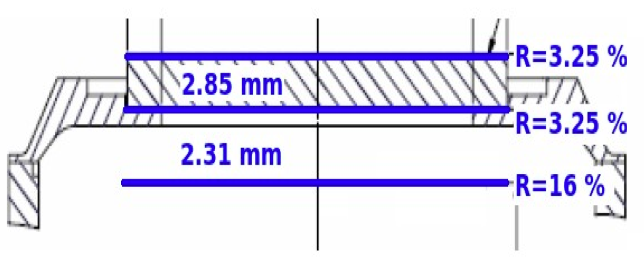}
\hfill
\includegraphics[width =0.4\textwidth, trim=6cm 0cm 6cm 1cm, clip=true]{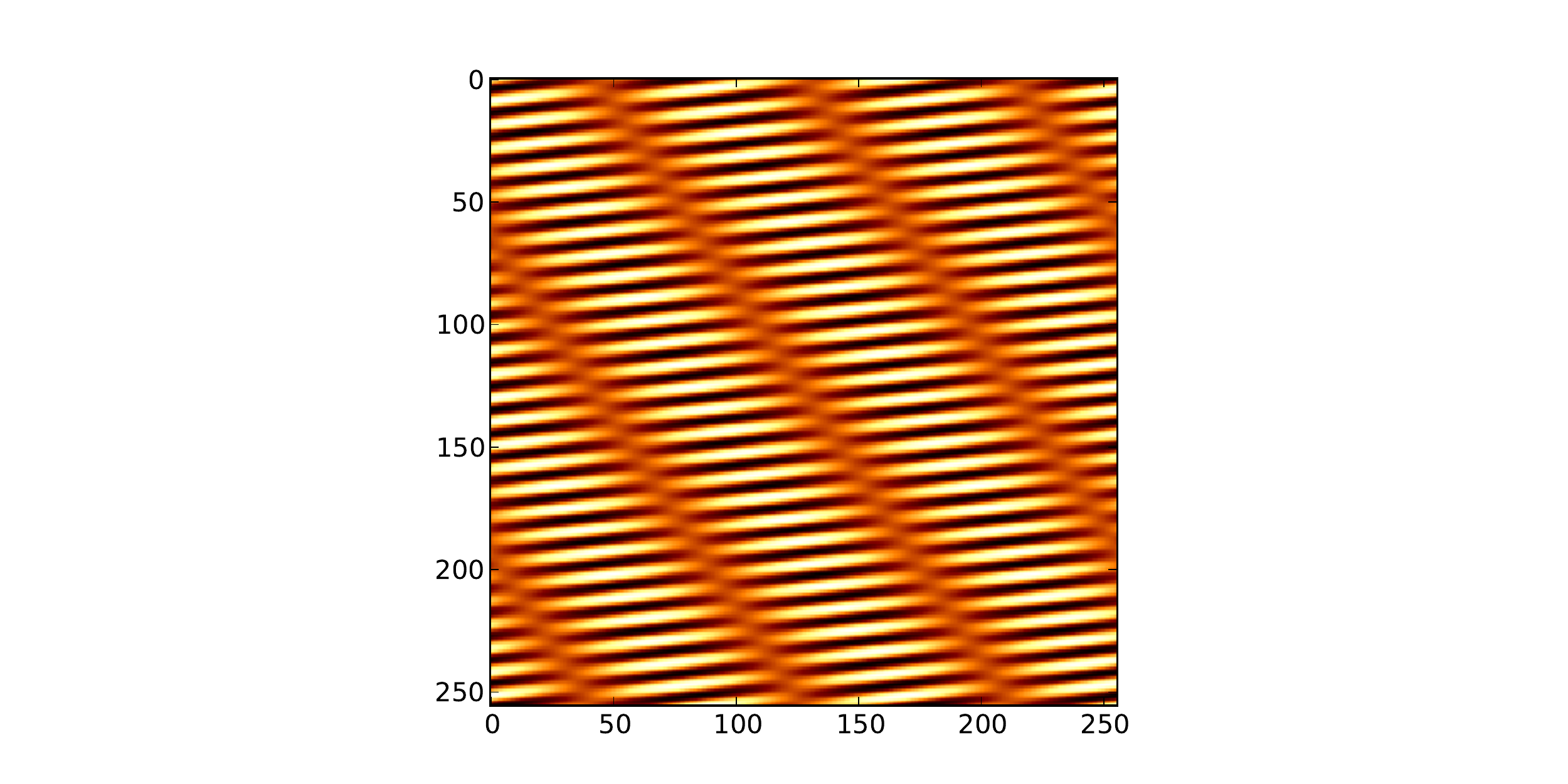}
\caption{Left: schematics of the IR camera leading to the fringing with the estimated distance and reflectivity of the different surfaces. Tilt angles between surfaces are $\sim 1.5$ arc-minute (not represented here). Right: simulated fringing when uniformly illuminating the camera with the metrology laser.$\alpha$ is here the angle between the camera detector and the plan defined by the exit splitter fibers.} \label{fig:camera_fringing}
\end{figure}

The camera used for the first tests was identified as a strong contributor to the measurement phase noise, as it suffered from a 3 components fringing between the detector, its thinned tilted substrate layer, and the entrance window of the camera. From a flat field image, we could estimate their properties (position, orientation, reflection coefficients), and model them properly to estimate their impact on the measured phases (Fig.~\ref{fig:camera_fringing}).
As we initially used the ABCD fringe coding from the tested IOBC, the fringes were sampled with only 4 pixels, distant by 5 to 6 pixels. The phase signal was therefore impacted by the fringing, and also highly dependent on the actual IOBC outputs position on the camera (in x,y and $\theta$). Fringing was also generating an artificial visibility change, but it is again difficult to disentangle it from, e.g., differential polarisation change between the splitter fibers\cite{perraut_1996a} .

\begin{figure}
\centering
\includegraphics[width=0.5\textwidth]{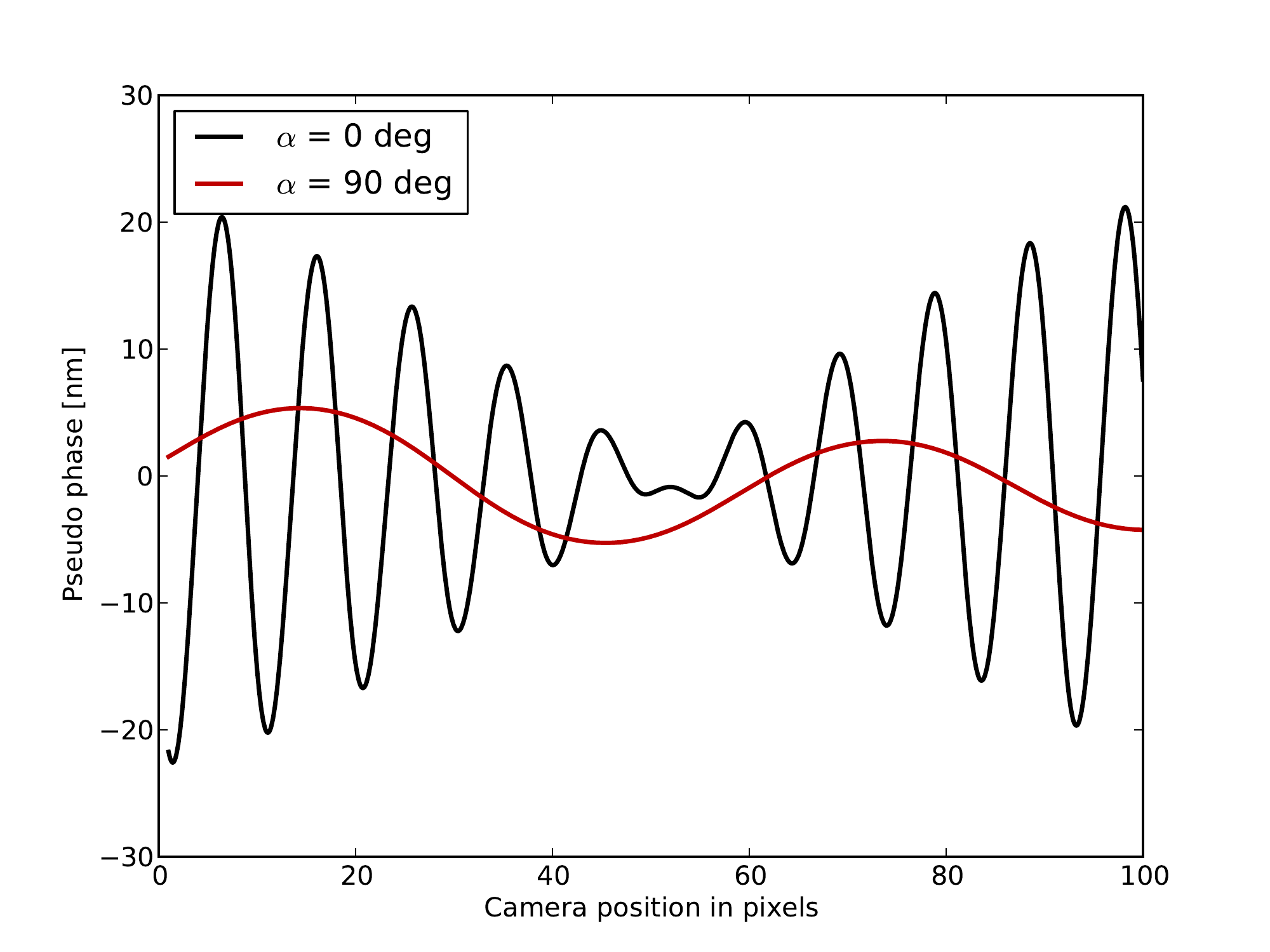}
\caption{Simulated effect of phase induced signal by the camera fringing only while moving the camera in front of the IOBC outputs.} \label{fig:phase_fringing}
\end{figure}

As shown in Fig.~\ref{fig:phase_fringing}, this generates a wrong phase signal with a few pixels frequency, and amplitude as high as 20 nm peak-to-peak. It is therefore difficult to assess the precision reached by these measurements, especially because the camera was oriented in the worst position ($\alpha = 0^\circ$).

In the modelling of the full system, we considered a linear displacement of the camera of a few pixels per Kelvin. The camera was oriented in the $\alpha = 0^\circ$ position of Fig.~\ref{fig:phase_fringing} corresponding to the measurement conditions.

     \subsection{Materials}

The material properties considered here are presented in Tab.~\ref{tab:materials}. They are all given at 25$^\circ$C. When data were not found, we used values of fused silica.
\begin{table} \centering
\begin{tabular}{lllllllp{0.1cm}l}
\hline \hline
     & $n_{\lambda=1908nm}$ & $\frac{dn}{dT}$ &  $\Delta n$ & $\frac{d\Delta n}{dT}$ & $\rho$ \\
     &                                         & [$K^{-1}$]         &                  & [$K^{-1}$]                 & [$K^{-1}$] \\
\hline
     Fused Silica        & 1.439 \:\:$^[$\cite{malitson_1965a,  kitamura_2007a}$^]$ & $8.1\cdot10^{-6} $ \:\:$^[$\cite{malitson_1965a,  kitamura_2007a}$^]$  &$2.8\cdot10^{-4}$ $^a$ & $5.7\cdot10^{-7}$ \:\:$^[$\cite{zhang_2011a}$^]$&$4.8\cdot 10^{-7}$ \:\:$^[$\cite{malitson_1965a,  kitamura_2007a}$^]$\\
     Exit splitter    & 1.48$^b$ & $1.1\cdot10^{-8}$ $^b$& $< 1.0\cdot 10^{-6}$ $^b$ &$1.0\cdot10^{-8}$ $^b$ &$8.2\cdot10^{-6}$ $^b$\\
     IOBC                   & 1.445 \:\:$^[$\cite{labeye_2008}$^]$&  $8.1\cdot10^{-6} $ $^{c}$ &   $10^{-4}$ \:\:$^[$\cite{jocou_2012a}$^]$ & $5.7\cdot10^{-7}$ $^c$ & $4.8\cdot 10^{-7}$ $^c$ \\
     \hline
\end{tabular}
\caption{Parameters of the different materials considered in the metrology modelling. Beating length and birefringence $\Delta n$ are used when intrinsic to the component (e.g. PANDA fiber). $\rho$ is the coefficient of thermal expansion. All parameters are given at 25$^\circ$C. Notes: $^a$ Fujikara PANDA PM1550 data sheet gives a beating length of 5.5\:mm at $\lambda=1550nm$; $^b$ Private com.; $^c$ No data available, fused silica value is considered. }\label{tab:materials}
\end{table}
%

\section{Comparison to tests}
\label{part:test_modeling}

We model now the test set-up as described in Sect.~\ref{part:testsetup} and \ref{part:met_modeling}. The goal is to assess its past performance when affected by camera fringing, and also to validate the model against reliable measurements.

	\subsection{Impact of detector fringing in past tests}

\begin{figure}[b]
\centering
\includegraphics[width=.5\textwidth, trim=2cm 0 19.3cm 0, clip=true]{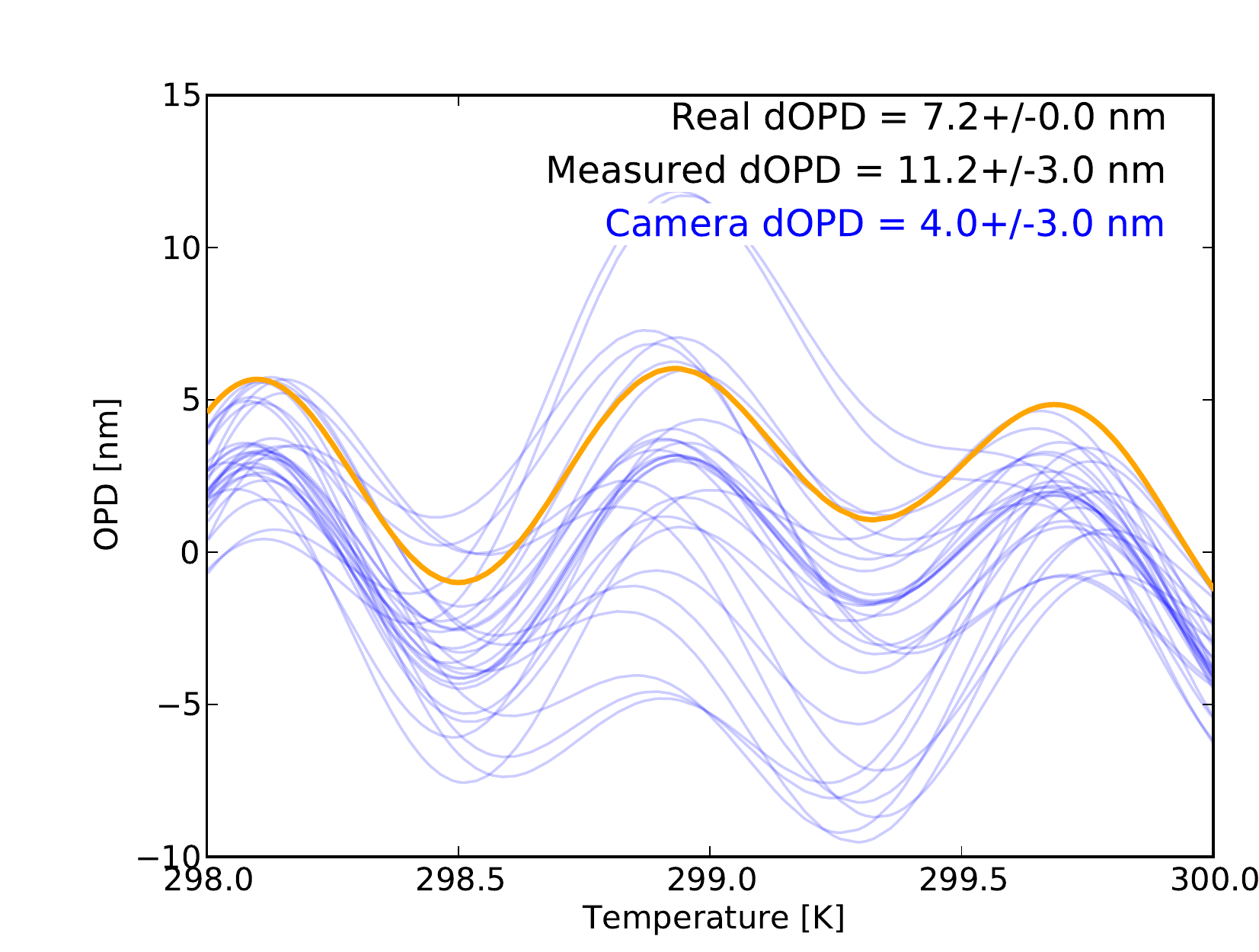}
\caption{Example of a test set-up simulation with SM fibers, and the camera at different positions. The yellow curve represents the real OPD generated by the set-up, while the blue curves show the {\it measured} OPD for different random positions of the camera when considering its fringing. On top right are computed the real OPD variation amplitude (peak-to-peak), the measured one and the dispersion for the different camera positions, and finally additional measured instability due the camera (i.e. difference between the measured and the real OPD).}\label{fig:camera_impact}
\end{figure}

	One of the main concerns of the previous tests was the impact of the camera fringing on the measurements. We made a series of simulations modeling the full metrology test set-up. For each simulation, the camera was positioned in the worst orientation in a random initial position, and was linearly moving in this direction at a rate of 5 pixel/K. The result of such a simulation is presented on Fig.~\ref{fig:camera_impact}. Overall the camera fringing makes the metrology look more unstable, by adding a peak-to-peak measurement instability of 5 $\pm$ 3\:nm depending on the camera position. Therefore, our past measurements using the IOBC as metrology phase analyser were not able to measure the metrology stability at a precision better than $\sim$10\:nm peak-to-peak. The dependence with the actual camera position can also explain the difficulty to reproduce results in the past.

	\subsection{Sensitivity to input polarisation}
	\label{part:pol_sensitivity}

\begin{figure}[t]
\centering
\includegraphics[width=0.45\textwidth]{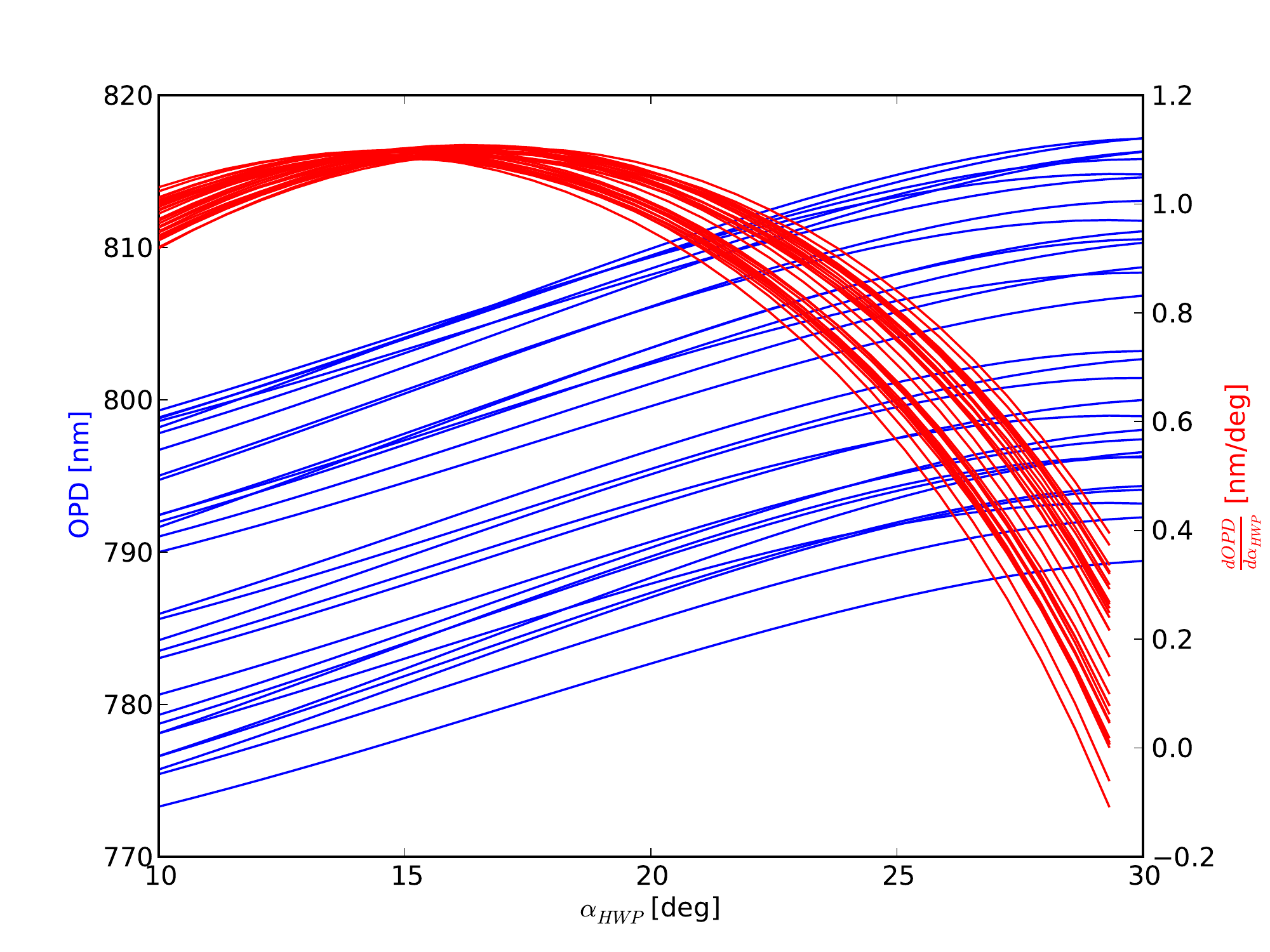}
\includegraphics[width=0.45\textwidth]{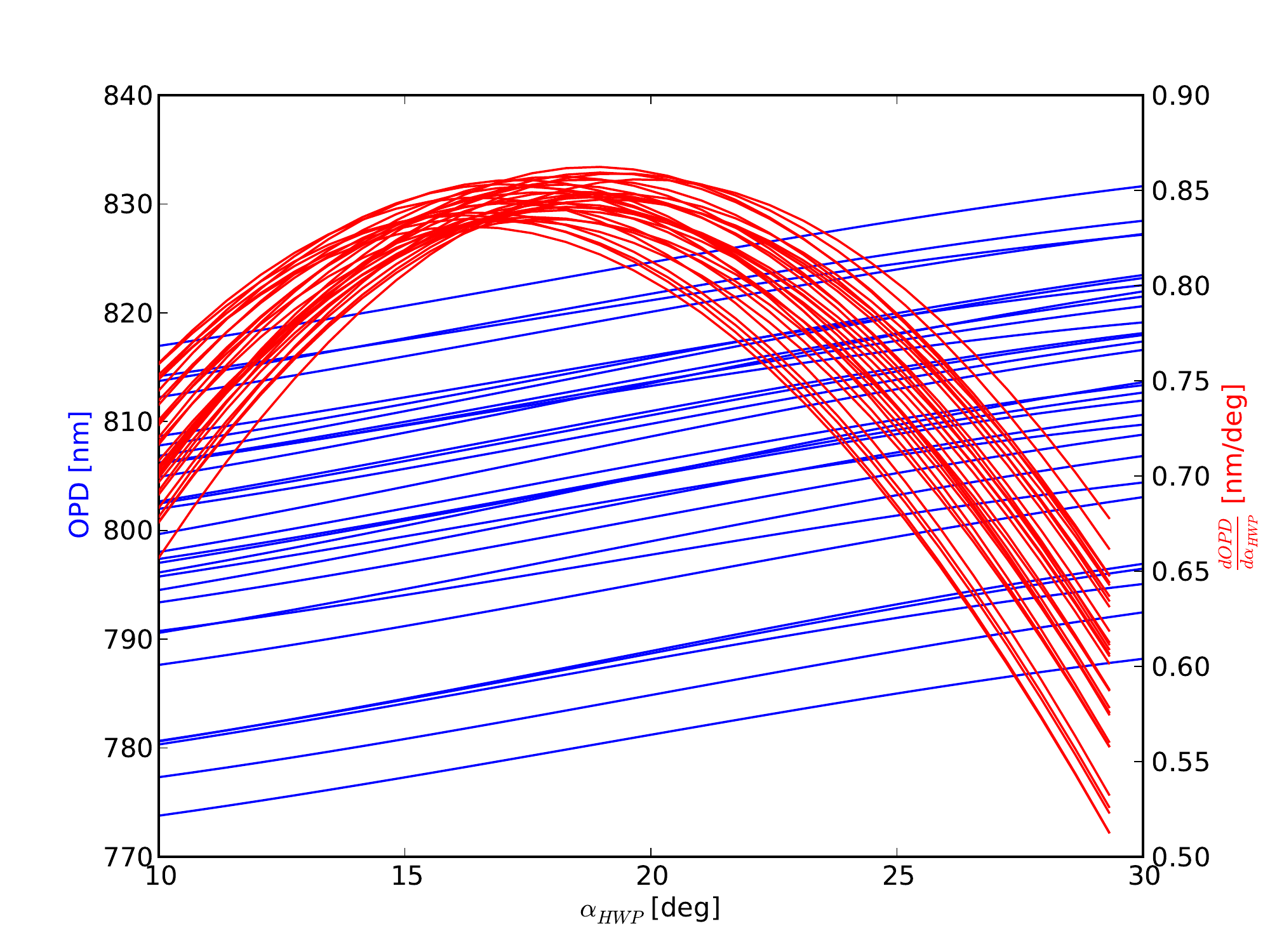}
\caption{Phase induced by rotating the input linear polarisation with an half-wave plate at an angle of 10$^\circ$ to 30$^\circ$. We present here the case of a SM fibers splitter (SM1 and SM2 angles of -40$^\circ$ and +20$^\circ$ respectively), a thermal sensitivity of 25K/W at  50mW power level. Left plot is with one entrance polariser, and the right one with two. Each curve correspond to a different set of random connectors.} \label{fig:pol_sensitivity}
\end{figure}
%


We checked the sensitivity of the injection set-up to input polarisation by actively rotating the linear polarisation with a half-wave plate in a fiber port before the first in-line polariser:
\begin{itemize}
     \item {\bf Baseline design}: with SM fibers the phase did depend on the input polarisation:
     \begin{itemize}
          \item Using 1 polariser in front of the splitter, behind the HWP: 1nm/deg at 20$^\circ$ angle.
          \item Using 2 polarisers in front of the splitter, behind the HWP: 0.5nm/deg at 20$^\circ$ angle.
          \item No test was done without polariser.
     \end{itemize}
	The test with output PM fibers and one entrance polariser was not conclusive.
     \item {\bf Alternative design}: for the alternative IOBC, the phase does not depend on the input polarisation, even in the setup with exit splitter still in. This is indicative of a low {\it differential} birefringence in the IOBC, in agreement with the conclusions of the various tests (e.g. see Sect.~\ref{part:heating_fiber_IO}).
\end{itemize}
No camera motion is considered considering the small time scale of these measurements (few minutes). Simulations show then that the camera fringing as an impact $\ll 0.1$\:nm/deg. We simulated the baseline design measurements: one or two in-line  polarisers before a SM splitter, and then a bulk polariser and the IOBC. Different fiber misalignments are tested, up to 60$^\circ$ from each other, worst case observed in the lab. Considering the alignment precision of the half-wave plate to $\pm2^\circ$, and the dispersion introduced by the connectors, the precision of the measurements is estimated to $\pm$0.2-0.3 nm/deg at a 20$^\circ$ angle. A set of results is summarized in Tab.~\ref{tab:pol_sensitivity}, and Fig.~\ref{fig:pol_sensitivity}. They can be considered in relatively good agreement with observations with one polariser at the entrance. When adding a second polariser at the entrance, the improvement is not as important as observed, but still within error bars considering the high number of free parameters, and the low number of measurements (one every 5 degrees).

Increasing the polariser extinction ratio did not reduce the sensitivity of simulations, as we are probably limited by FC connectors (ER$\sim$15, against ER=23 for in-line polarisers). The sensitivity to input polarisation is also dependent on heat (i.e. laser power, splitting ratio or heating coefficient), and does not show a linear behavior in this respect. For a null power, we still observe up to 0.3\:nm/deg with one polariser, which then results from the various misalignments. Adding a second polariser reduces dramatically the sensitivity by a factor of 100.

\begin{table}[b]
\centering
\vspace{1em}
\begin{tabular}{l|c|cccc}
\hline \hline
 & Measurements &$\alpha_{SM1}=$ 0$^\circ$ & 0$^\circ$ & -20$^\circ$ & -40 $^\circ$ \\
& &$\alpha_{SM2}=$ 0$^\circ$ & 20$^\circ$ & +20$^\circ$ & +20$^\circ$ \\
\hline
1x polariser & 1 nm/deg &0.23 nm/deg & 0.3 nm/deg & 0.85 nm/deg & 1.0 nm/deg \\
2x polarisers & 0.5 nm/deg & 0.15 nm/deg & 0.3 nm /deg & 0.65 nm/deg & 0.85 nm/deg \\
\hline
\end{tabular}
\vspace{0.5em}
\caption{Input polarisation sensitivty of the model for an entrance power of 50\:mW and different configurations of the exit splitter SM fibers.}\label{tab:pol_sensitivity}
\end{table}

	\subsection{Power sensitivity}
\label{part:power_sensitivity}

The power sensitivity of the Fizeau test set-up was recently estimated by monitoring the phase during power cycles between 100 and 400\:mW. We recently measured a power sensitivity of the order of 1.2-1.6\:nm/mW for the baseline design and a SM exit splitter, and an almost negligible sensitivity for the alternative design ($<0.1$\:nm/mW)\cite{kok_2014a}. This result emphasizes again the importance of differential heating in the baseline design, and is therefore dependent on the splitting ratio of the exit splitter. For the alternative design, the phase effect is mainly due to the combined effect of birefringence, splitter/IOBC misalignment and IOBC asymmetries, if not simply due to mechanical instability at such a low level. Tab.~\ref{tab:power_sensitivity_simu} presents the result of baseline and back-up design simulations when changing the splitters splitting ratio (exit splitter and IOBC), and for heating of 25\:K/W. Even in the worst case, we miss a factor 2 to 3 compared to measurements, which could be due to: {\it 1)} an underestimated laser heating; {\it 2)} thermal expansion of the fiber holder, which can still reach a nanometer level, even with a thermal control to $\pm$\:3\:mK.

\begin{table}[b]
\centering
\begin{tabular}{lcccc}
\hline \hline
Splitting ratio & 1.05& 1.1 & 1.2 & 1.4 \\
Splitting & 49/51 & 47/53  & 45/55 & 42/58 \\
\hline
Baseline  [nm/mW]  & 0.315 & 0.360 & 0.455 & 0.610\\
Back-up [nm/mW] & 0.002 & 0.009 & 0.014 & 0.022\\
\hline
\end{tabular}
\vspace{0.5em}
\caption{Simulated power sensitivity of the baseline design (with SM splitter; $\alpha_{SM1}=-40^\circ$, $\alpha_{SM2}=+20^\circ$), and alternative design ($\alpha_{SM}=+20^\circ$). The splitting ratio is given for the exit splitter in the baseline case, and for the first coupler in the IOBC for the alternative one.} \label{tab:power_sensitivity_simu}
\end{table}

%
%
%
%

\section{Metrology performance analysis}
\label{part:fullmet_modeling}

We finally model one injection unit in the VLTI environment, i.e. for a daily ambient temperature variation of $\sim$100\:mK. Since the "zero point" of the model depends on various parameters, we consider a wider range and study 100\:mK sections. 
The daily temperature variation will impact the 30\:m long PM fiber that links the metrology laser (in the Combined Cou\'e room) to the GRAVITY cryostat (in VLTI lab), and should lead to higher sensitivity of the metrology than observed in the lab with the 2-m long fiber. Note that we only look at the stability due to the polarisation change generating up to 5\% of power fluctuations, while the laser power is generally stable to 0.5\%.

In practice the temperature of the splitter and IOBC was controlled to $\pm$\:3\:mK. The temperature control loop of the splitter and IOBC was simulated by reducing the heating effect to 2 to 5\:K/W so as to maintain their temperature in this range. This is however a clearly unrealistic model since it limits temperature variations, but cannot mimic the real behavior of the controller with time (which could lead to additional instabilities in stable periods because of noise, and vice versa). 
Also, the temperature controllers and sensors are not in direct contact with the components surface, so the waveguide temperature is probably less stable and suffer from higher temperature change than measured. We should then see the results in this situation as a best case. Considering an heating of 25\:K/W (no cooling), the OPD variations can then increase by a factor of 10 or more.

\begin{figure}
\centering
\includegraphics[width=0.45\textwidth, trim=2cm 0 19.1cm 0, clip=true]{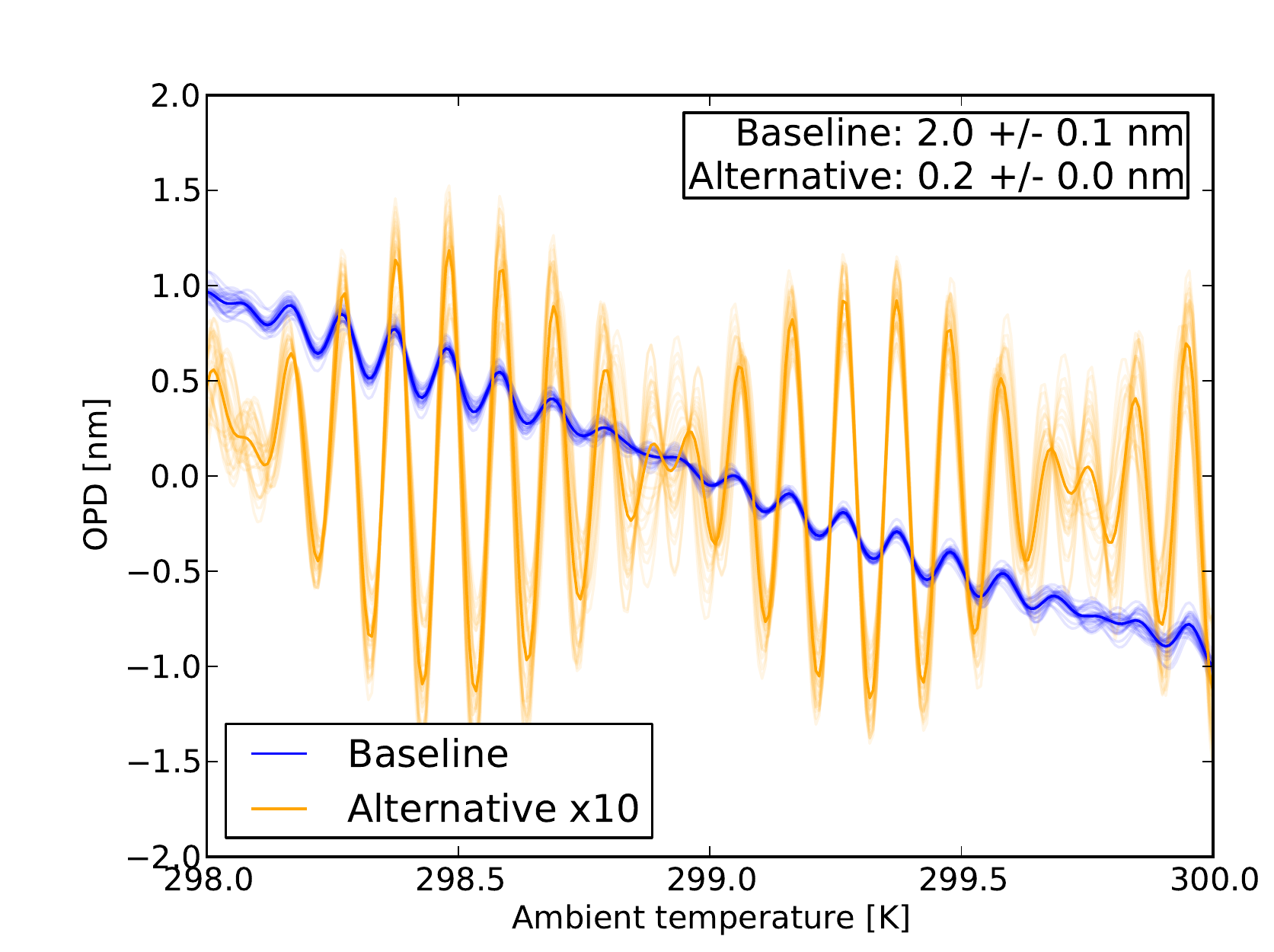}
\includegraphics[width=0.45\textwidth, trim=2cm 0 19.1cm 0, clip=true]{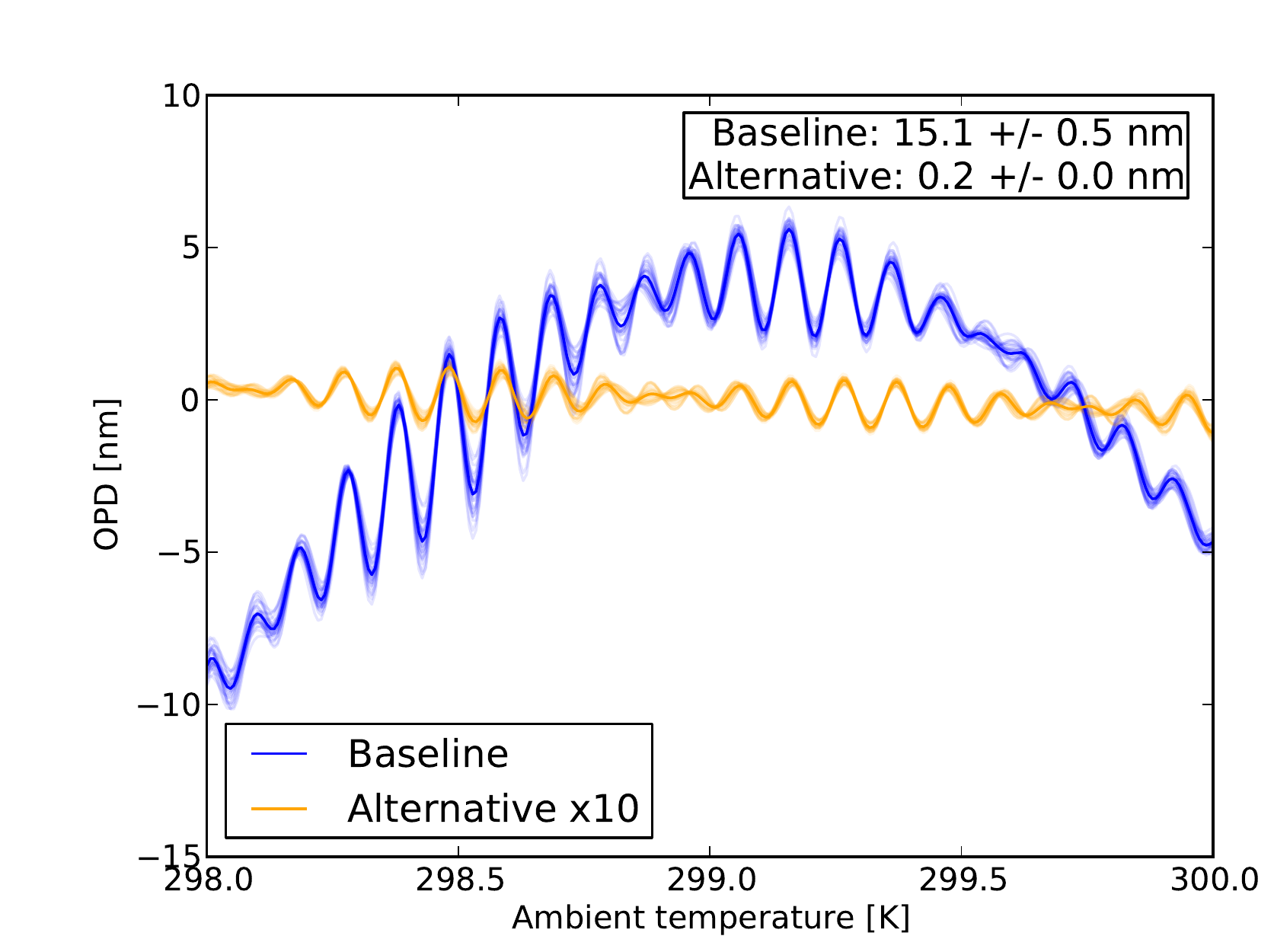}\\
\includegraphics[width=0.45\textwidth, trim=2cm 0 19.3cm 0, clip=true]{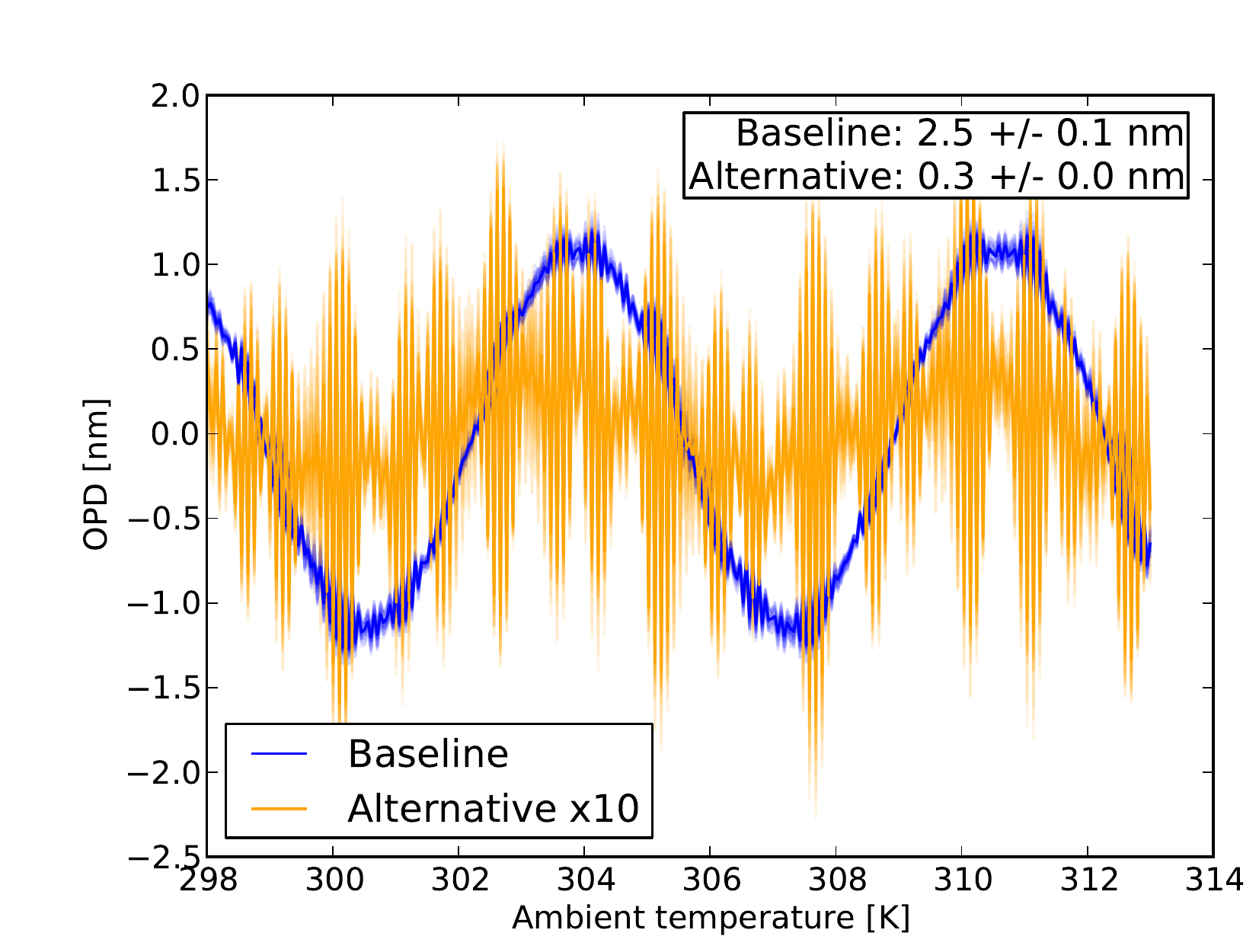}
\includegraphics[width=0.45\textwidth, trim=2cm 0 19.3cm 0, clip=true]{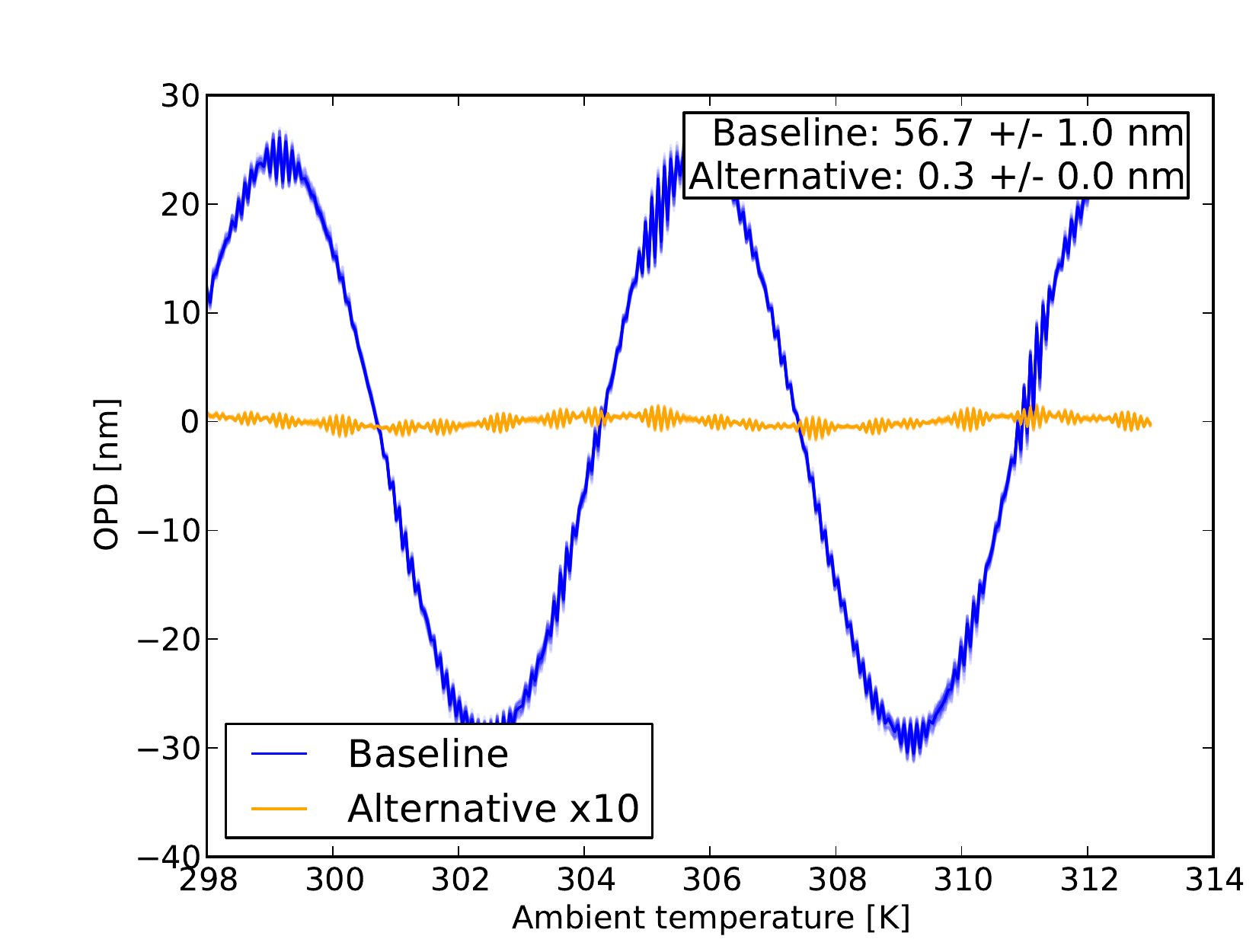}\\
\hrule
\includegraphics[width=0.45\textwidth, trim=2cm 0 19.3cm 0, clip=true]{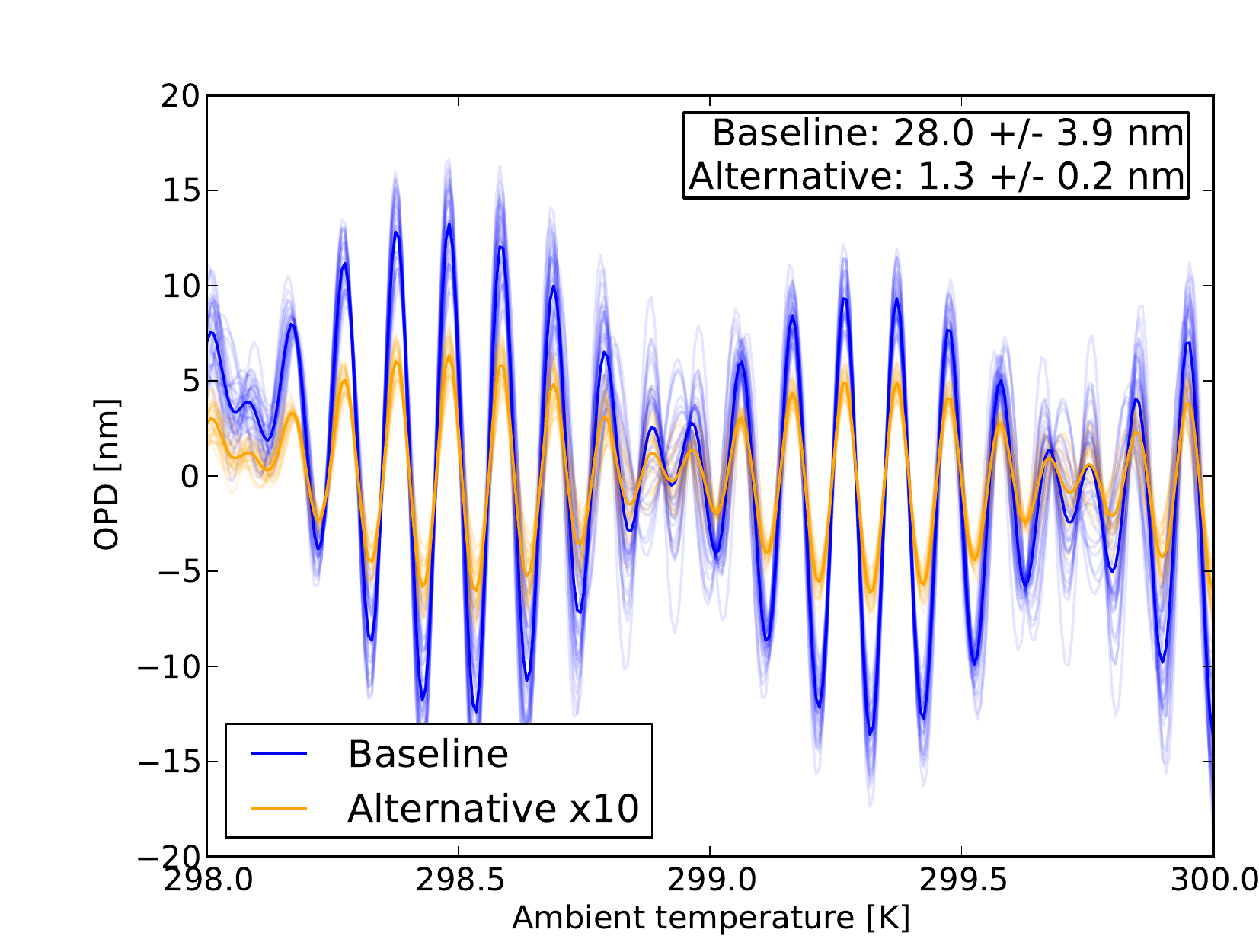}
\includegraphics[width=0.45\textwidth, trim=2cm 0 19.3cm 0, clip=true]{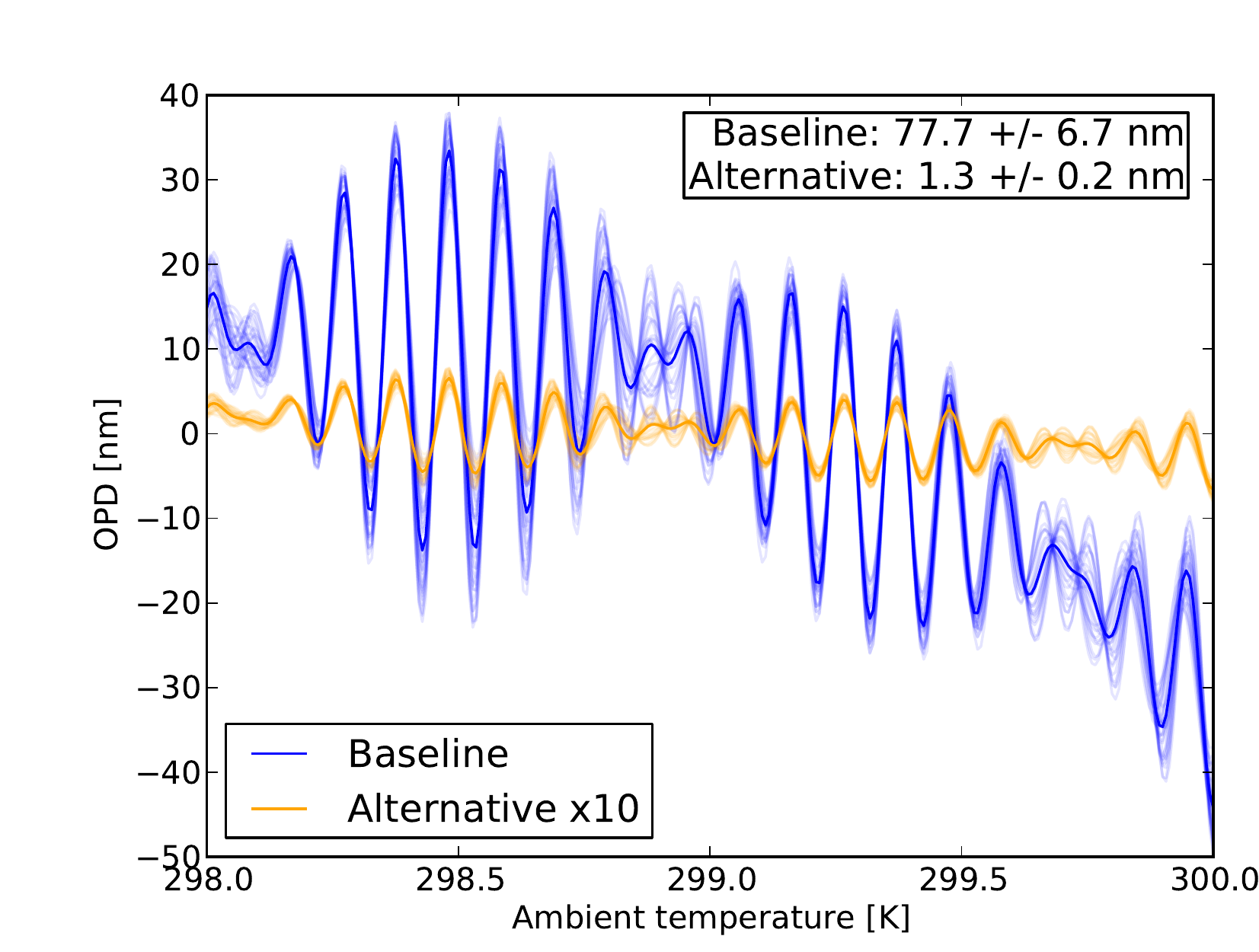}
\caption{Results in "VLTI conditions" with a 30-m long PM fiber for a PM and SM splitter (left and right respectively), with splitter and IOBC temperature control on top ($\Delta T = 2$\:K) and center ($\Delta T = 15$\:K), and no thermal control on bottom. Blue curves represent the baseline design, and yellow the alternative one (x10 for sake of clarity). Each dim curve represents a different set of connectors, and the text box shows the average and dispersion of the phase peak-to-peak amplitude for 50 different sets of connectors. The splitter splitting ratio is set to 1.4 in all cases.}\label{fig:vlti_conditions}
\end{figure}

Fig.~\ref{fig:vlti_conditions} shows the result of different simulations for an exit splitter with PM or SM fibers. For the PM case, one fiber is aligned to the IOBC axes, and the other is misaligned with an angle of 2$^\circ$. For the SM case, both fibers are misaligned by +20$^\circ$ and -40$^\circ$, as we could measure. Simulations with random splitter/IOBC misalignments show a small impact of the fiber misalignment in the PM case, and that we are close from worst case in the SM one. Because of the limitations of the model, the metrology phase instability also shows a 10-20\:K periodicity. These simulations show two regimes:
\begin{itemize}
\item A very sensitive variation of up to 1\:nm/mK due to the long entrance PM fiber generating laser power fluctuations;
\item A slower "drift" (on a small temperature scale) of up to few 10\:nm/K, resulting from the exit splitter and IOBC misalignments and asymmetries.
\end{itemize}
%
%

	\subsection{Impact of connectors}

The different components being birefringent, it is important to keep their axes aligned. The impact of FC/APC connectors and splices is studied with random sets of connectors. It appears they have little impact on the phase amplitude variation. Although temperature sensitivity can change by 50\% depending on the set of connector, the effect is localised to a limited range of temperature of a few 10\:mK maximum, thus not generating a significant additional phase effect. We are therefore limited by the exit splitter and IOBC, and splicing the components in the cryostat does not appear as necessary.

	\subsection{Impact of exit splitter fibers and polariser for the baseline design}
	
Past measurements seem to indicate a better stability of the SM splitter, believed to originate from their lower birefringence. Simulations on the other hand suggest a significantly better stability of the PM exit splitter (2\:nm peak-to-peak) compared to the SM one (15\:nm on 2\:K range, 57\:nm peak-to-peak), because of the better polarisation alignment that can be achieved with PM fiber. Achieving an alignment to $\pm 2^\circ$ with SM fibers would dramatically reduce the phase instability to 2\:nm peak-to-peak, equivalent to the PM solution. 
Using directly an integrated Y-junction matching the IOBC outputs as splitter (without fibers) could allow an even cleaner injection in the IOBC.

The bulk polariser also creates differential heating in the IOBC because of polarisation rotation in the splitter. However the induced phase effect is negligible once compared to the case where the bulk polariser is removed, and where both polarisations are mixed together in the IOBC and suffer from its strong birefringence.

	\subsection{Baseline vs alternative design}
			
As expected, the alternative design shows an excellent stability with respect to ambient temperature variations and power. Although we miss numbers for a precise model and can consider it as optimistic, it would be surprising it is less stable than the baseline design because of the exit splitter effect. The caveat of the model resides in the Y-junction which is reached by a multi-mode beam. Although we measured its phase sensitivity depending on beam injection quality, we do not know its behavior concerning splitting ratio, which could then be another source of instability as seen in Sect.~\ref{part:power_sensitivity}.

The baseline design on the other hand shows a very high sensitivity with ambient temperature because of the entrance PM fiber: on scales of only 50\:mK, the measured OPD can change by several nanometers. This is especially true with a SM splitter, although a PM one show a similar behavior at a lower degree.

The drawback of the alternative design is to lose 2 out of the 24 IOBC outputs, then reducing the instrument transmission by 8\%, and, more importantly, decreasing the Fringe Tracker limit magnitude by Kmag $\sim$0.5. An alternative could then be to implement the baseline on the Fringe Tracker side to allow the instrument observing the faintest sources possible in stable conditions, while implementing the alternative design on the science spectrometer. The metrology would then suffer from non common path error on one spectrometer only.

	\subsection{Unpolarised laser}	
	\label{part:unpol_laser}

By using an unpolarised laser, the fiber effect before the entrance polariser in the cryostat is nulled and the injection unit is only subject to the laser power fluctuations. The OPD instability would be less than 0.040\:nm/mW and 0.005\:nm/mW for the baseline (highly misaligned) and alternative designs according to simulations. Lab measurement have demonstrated a laser power stability of 0.5 to 5\% RMS at 2\:W power level, that is 0.4 to 4\:nm instability for the baseline design. The laser power dissipation in the polariser must however be studied.

\section{CONCLUSIONS}
\label{part:ccl}

We presented a set of simulations aiming at estimating the performance and understanding the behavior of the GRAVITY metrology. From the lessons learned with this model, we can think of several improvements of the current metrology set-up:
\begin{itemize}
\item The main source of instability is the linearly polarised laser and the 30-m long PM fiber. A first improvement would be to use a PM fiber with a better Extinction ratio, so as to minimize the power fluctuations at the entrance polariser level. A second solution could be to use an unpolarised laser as explained in Sect.~\ref{part:unpol_laser}, although the effect of power dissipation has to be studied.
\item The model and lab measurements emphasize that the baseline design instabilities arises from the differential heating originating in the unbalanced exit splitter in front of the IOBC.
\item Although the alternative design shows a better stability, it involves losing 2 out of the 24 IOBC outputs. This is impacting the Fringe Tracker performance by making it loses roughly 0.5 mag sensitivity (private comm.).
\item Simulations and measurements seem to disagree concerning the "superiority" of SM fibers for the exit splitter. Nevertheless, special effort should be put in aligning properly the SM fibers during the gluing phase, in a fiber configuration as close as possible from the cryostat one. According to our understanding of fiber components, injecting the metrology in the IOBC directly via an integrated Y-junction (without fibers) is a promising solution.
\item The impact of the entrance splitter that splits the metrology light between the two spectrometers was not included in these simulations. As it was observed, its splitting ratio varies with input power, it would be valuable to exchange it for an integrated version that demonstrated constant properties up to $P_{laser}$=1\:W.
\end{itemize}

The model suffers from few caveats making difficult to assess the real metrology performance from modeling only, although simulations and measurements agree to a reasonable level. The main caveats are:
\begin{itemize}
\item The number of components and free parameters is large. It is possible to assess the impact of certain unknown parameters like connectors, but additional characterisations would be necessary for critical components like the exit splitters or the IOBC.
\item A realistic model of fibered component is difficult since their behavior highly depends on the real routing (i.e. mechanical stress). This is especially true for SM fibers, but hopefully their number is limited, and their routing and behavior are known to a reasonable level.
\item A complete metrology model must take into account numerous physical processes: polarised light propagation, thermal stress, mechanical constraints, natural and active cooling/heating, few modes waveguide behavior, etc.
\end{itemize}
Given all these constraints, the model reproduces observations at an acceptable level. Improving the model can be done by implementing a realistic thermal control loop and laser power fluctuations, and a more complete characterisation of the different fiber components. For instance, the entrance splitter shows a splitting ratio varying with the laser power, which cannot be explained with the current, standard model.

\bibliographystyle{spiebib}
\bibliography{metrology_modeling_v1}

\end{document}